\documentclass[a4paper,11pt]{article}
\pdfoutput=1 
\usepackage{jheppub} 
\preprint{TIFR/TH/26-8}
 \usepackage{graphicx}
 \usepackage{subfig}
 \usepackage{caption}
 \usepackage{subcaption} 
\graphicspath{ {./} }                    
\usepackage{slashed}
\usepackage[T1]{fontenc}
\usepackage{amssymb}
\usepackage{mathrsfs}
\usepackage{amsmath}
\usepackage{dsfont}
\usepackage{float}
\usepackage{hyperref}
\usepackage{cancel}
\usepackage[normalem]{ulem}
\usepackage{xcolor}
\usepackage{color}
\usepackage{braket}
\usepackage{bm}
\usepackage{calligra}
\usepackage{feynmp}
\usepackage{tikz}
\usepackage{tikz-feynman}
\usepackage{hyperref}
\usepackage{orcidlink}
\tikzfeynmanset{compat=1.1.0}
\usetikzlibrary{positioning}
\newcommand{\beq}{\begin{equation}}
\newcommand{\eeq}{\end{equation}}
\def\barr{\begin{array}}
\def\earr{\end{array}}

\newcommand{\bsp}{\begin{split}}
\newcommand{\esp}{\end{split}}
\newcommand{\bit}{\begin{itemize}}
\newcommand{\eit}{\end{itemize}}
\definecolor{darkcyan}{cmyk}{1,0,0,0.4}
\definecolor{darkgreen}{cmyk}{1,0,1,0.4}

\def\lapp{\mathrel{\rlap{\raise.5ex\hbox{$<$}}
                    {\lower.5ex\hbox{$\sim$}}}}
\def\gapp{\mathrel{\rlap{\raise.5ex\hbox{$>$}}
                    {\lower.5ex\hbox{$\sim$}}}}
\begin{document}

\title{\boldmath Return of the CHAMPs: A clockwork portal to charged dark matter}

\author[a]{Debajyoti Choudhury \orcidlink{0000-0002-8124-0043},}
\emailAdd{debchou.physics@gmail.com}
\author[a]{Vineet K. Jha$^{\dagger}$ \orcidlink{0009-0000-2605-2544},}
\emailAdd{vineet.phd2022@physics.du.ac.in}
\affiliation[a]{Department of Physics and Astrophysics, University of Delhi,
  Delhi 110 007, India.}
\author[b]{Suvam Maharana \orcidlink{0000-0001-9313-4064}}
\emailAdd{suvam.maharana\_528@tifr.res.in}
\affiliation[b]{Department of Theoretical Physics, Tata Institute of Fundamental Research, Homi Bhabha Road, Mumbai 400005, India.}

\abstract{{\def\thefootnote{$\dagger$}
\footnotetext{Corresponding author.}}\hspace{-0.66em}While Dark Matter (DM) is conventionally assumed to be
  chargeless, the possibility of a charged massive particle (CHAMP) as
  the DM particle remains alive. With phenomenological constraints on
  the charge being very severe, such a scenario is
  often sought to be dismissed, citing naturalness. Moreover, the establishment of the correct relic density is often a concern. We demonstrate
  here that such a (mini)charged DM can yet be realized within the
  clockwork paradigm, without the need to invoke unnaturally small
  parameters. The model is examined against constraints, theoretical and experimental, and the phenomenologically admissible parameter space is delineated. Several intriguing tests, at the LHC as well as at future direct and indirect detection experiments, are pointed out.}
\maketitle
\flushbottom
\section{Introduction} 
\label{sec:intro}

The nature of dark matter (DM) in the observable Universe still
remains frustratingly obscure. While its existence (and its
overwhelming contribution to the matter-energy density of the
Universe) can be inferred from its gravitational interactions, very
little is known about the properties of the DM particle, whether it be
its mass or spin, or its participation in the other known fundamental
interactions.  The standard strong interactions are disfavoured, as
that would lead to stable exotic bound states, and no such states have
been seen. Similarly, strong self-interactions would, nominally, lead
to significant pressure for a DM cloud, thereby interfering with
large-scale structure formation. Nonetheless, certain assertions with
regards the DM's mass and its possible interactions with the visible
sector are often made from a purely phenomenological point of
view. Although these assumptions may not necessarily have a strong
theoretical or observational motivation, they appear to be the most
natural choices within minimal extensions of the Standard Model (SM)
and, therefore, prove to be essential testing grounds for DM
model-building. One such assertion is that of a weak-scale [$ \sim
  \mathcal{O}(100 \, \mbox{GeV} - 1 \, \mbox{TeV})$] DM, a natural
possibility considering that the fundamental scale of the visible
sector is \emph{presumably} set by electroweak symmetry breaking (EWSB). Of
course, this is already otherwise motivated from the fact that a
weak-scale DM interacting with visible matter through a weak-strength
coupling automatically produces the correct thermal relic abundance in
the Universe --- the so-called WIMP miracle. Then, for the simplest
case of a weak-isospin singlet DM, minimalism dictates that
nongravitational DM-SM interactions can exist either through the Higgs
portal or through the $Z$ portal. Note, though, that if the DM is
indeed a SM-singlet, the gauge-invariant completion of the required
DM-$Z$ interactions can only be furnished by higher-dimensional
operators involving DM couplings with the Higgs doublet and its
covariant derivative \cite{DeSimone:2014qkh, Arcadi:2014lta}. However,
the DM-SM couplings dictating the thermal freeze-out relic as well as
the DM-nucleon scatterings in both these minimal portals have largely
been ruled out for a weak-scale DM by negative results in direct
detection experiments\footnote{Assuming a nonstandard cosmology,
however, may still render a region of the parameter space of these
models viable \cite{Chanda:2019xyl}.} (see \emph{e.g.}
ref.\cite{Escudero:2016gzx}).

Although relatively less DMexplored
for a heavy relic, it is possible in general that it carries a
nonzero electric charge, most naturally through a nonzero
hypercharge. Such electrically CHArged Massive Particles (CHAMPs) were
originally proposed in ref.\cite{DeRujula:1989fe} with the assumption
that they carry unit or integral charges and were subsequently
reexamined in ref.~\cite{Dobroliubov:1989mr} for fractional charges as
well. Expectedly, CHAMP candidates with integral charges were soon
ruled out by the nonobservation of heavy exotic atoms as well as from
direct searches \cite{Dimopoulos:1989hk}.

Studied extensively for light DM candidates (see \emph{e.g.} refs. \cite{PhysRevLett.122.071801, Haas:2014dda, MiniBooNE:2018esg, Kelly:2018brz}), fractionally charged particles (with the charge being parametrized as $Q_{DM}=\epsilon e$), too, are subject to strong cosmological, astrophysical, and terrestrial constraints \cite{Davidson:2000hf}. For instance, the suppression of invisible decay modes of the ortho-positronium requires \( \epsilon \lesssim
10^{-5} \) for \( m_{DM} < m_e \)~\cite{PhysRevD.75.032004}. Similarly, the SLAC accelerator
experiment places upper limits of \( \epsilon
\lesssim 4.1 \times 10^{-5} \) for particles with mass around 1~MeV
and \( \epsilon \lesssim 5.8 \times 10^{-4} \) for particles with mass
around 100~MeV~\cite{PhysRevLett.81.1175}. Stellar constraints are
particularly stringent for \( m_{DM} \lesssim 10~\text{keV} \), from
plasmon decays into such particles, with the strongest bounds
($\epsilon \lesssim 10^{-14}$) emanating from observations of low-mass
red giants and white dwarfs~\cite{Davidson:2000hf,
  Raffelt:1996wa}. Analogously, if they were relativistic during the
epoch of big bang nucleosynthesis (BBN)---{\em i.e.,} $m_{DM} \lesssim {\cal
  O}({\rm MeV})$--- they would impact the expansion rate of the
universe during BBN, and hence the synthesis of light elements (while
the effective number of relativistic species is predicted to be \(
N_{\text{eff}} = 3.046 \) within the SM, current Planck results
constrain it to \( N_{\text{eff}} = 2.99 \pm 0.17
\)~\cite{Planck:2018vyg}).

It was argued in ref.\cite{Chuzhoy:2008zy} that the constraints
emanating from terrestrial experiments may be relaxed by exploiting
the possibility that the halo CHAMPs are either stopped from
penetrating the galactic disk by its magnetic field or are expelled
due to interactions with supernovae remnants\footnote{Note that such DM expulsions from the Galactic Disk are generally expected to come in conflict with estimates of the local DM density \cite{Bovy:2012tw,2018Gaia,Evans:2018bqy}, especially when the relic particle constitutes the entirety of the cosmological DM abundance \cite{Munoz:2018pzp}.}, provided the CHAMP
masses fall in the range $100 (Q_{DM}/e)^2~{\rm TeV}\lesssim m_{DM} \lesssim 10^8 (Q_{DM}/e)$ TeV. Consequently, a heavy CHAMP possessing a tiny $Q_{DM}$ is not subject to such expulsions, and for the scenario where it saturates the DM abundance, the present-day direct detection experiments place the most stringent upper bounds on its electric charge, namely $Q_{DM} \lesssim 10^{-10} e$, for $m_{DM} \lesssim 10^5$ GeV \cite{Dunsky:2018mqs, ParticleDataGroup:2024cfk, McDermott:2010pa}.
  
From a model-building perspective, then, it becomes a challenge to
naturally invoke such a tiny charge for the CHAMP. A
quasi-phenomenological route would be to posit a dark sector composed
of particles charged under a new Abelian gauge group $U(1)_C$. If the
corresponding gauge boson (the dark photon $C_\mu$) kinetically mixes
with the SM hypercharge boson ($B_{\mu}$) through a term $\epsilon
B_{\mu\nu} C^{\mu\nu}$, it can result in the new particle(s) acquiring
an effective electric charge proportional to the mixing parameter
$\epsilon$. The smallness of $\epsilon$ is sought to be explained by
requiring it to appear only through radiative corrections, to be
engendered by introducing particles that carry both $U(1)_Y$ and
$U(1)_C$ charges. However, restricting $Q_{DM} \lesssim 10^{-10} e$
requires a large suppression in the kinetic mixing, which can be
realized if it is induced only at an adequately high loop order \cite{Gherghetta:2019coi}, typically necessitating the introduction of new symmetries and/or multiple new fields\footnote{A kinetic mixing of such a small magnitude could possibly be induced within string theoretic/extra-dimensional setups as well, see \emph{e.g} refs.\cite{Abel:2003ue, Abel:2008ai, Goodsell:2009xc}. However, from an effective four-dimensional point of view, these too, in some form, would appear to be elaborate constructions comprising multiple new degrees of freedom (at least a spectrum of gravitons).}. Positing such a tiny charge, typically, also brings into question the standard means of ensuring the correct relic abundance.

In this work, we construct, instead, a model wherein the tiny charge
is generated via a localization of the SM photon in a theory
space. Appealing to the clockwork (CW) paradigm \cite{Choi:2014rja,
  Kaplan:2015fuy, Choi:2015fiu, Giudice:2016yja}, we enlarge the
Abelian factor of the SM gauge group as $U(1)_Y \to U(1)^{N+1}$. A
spontaneous breaking of the symmetry at a scale $f \gg m_{EW}$ by $N$
\emph{link} scalars $\Phi_j$ (with unequal charges $Q_{j+1}$ and $Q_j$
under adjacent $U(1)$'s) then leads to a theory of Abelian gauge
bosons with nearest-neighbour mass terms. With one subgroup
$U(1)_{CW}$ remaining unbroken, the vector boson spectrum contains a
massless particle along with a tower of heavy partners with masses
$\sim g q f$, where $g$ denotes the universal gauge coupling of the
$U(1)$'s and $q$ defines the ratio $|Q_{j+1}/Q_j|$. As long as
$q \gtrsim 2$, the
clockwork mechanism dictates that the massless vector field, which can
be identified with the SM hypercharge field $B_{\mu}$, is localized near
the $0$-th site on the theory space lattice. Consequently, if the
CHAMP $\chi$ is assumed to be charged only under $U(1)_N$, it would
possess a coupling with $B_{\mu}$ that is suppressed by an exponential
factor $\sim q^{-N}$. On the other hand, if the SM fermions and the
Higgs field are charged under $U(1)_0$ with the same quantum numbers
as the standard hypercharge assignments, their coupling with $B_{\mu}$
can, in principle, be commensurate with the SM values.

Understandably, phenomenological consistency constrains the \emph{a
priori} free parameters of the clockwork sector, namely $q$, $N$ and
$g$. For instance, the experimental constraint on the CHAMP's electric
charge correlates $q$ and $N$, whereas the value of the gauge coupling
$g$ (chosen, {\em for simplicity,} to have a uniform value over the
lattice) is fixed by demanding that the photon and the $Z$ couplings
to visible matter are precisely SM-like. The spontaneous symmetry breaking (SSB) scale $f$, which also
specifies the masses of the heavy CW vector bosons
(the harbingers of a set of $Z'$s) as well as the link scalars, is
bounded from below by the current collider limits on the masses of
neutral gauge bosons as well as the electroweak precision
observables. Interestingly, the $Z'$s play a pivotal role in setting
the relic abundance of the CHAMP DM candidate (which remains a
  problem in a stand-alone CHAMP theory with such a small $\epsilon$) through the freeze-out mechanism. We find that in our
minimal and constrained setup, with the CHAMP being a Dirac fermion,
the primary annihilation channel for $\chi$ is $\chi \Bar{\chi} \to f
\Bar{f}$, where $f$ denotes the SM fermions, and mediated by the
$Z'$s. The correct relic abundance is most easily obtained when
$m_{\chi} \sim m_{Z'}/2$. Since electroweak precision and the
prevailing collider constraints push the lower limit on the $Z'$ mass
scale in the model to the multi-TeV range, the allowed DM mass is also
accordingly restricted to be around or above the TeV scale.  Thus, for
$f$ near the TeV scale, the model naturally accommodates a minicharged
CHAMP DM of mass in the range $\sim 0.5 - 1$ TeV with $\mathcal{O}(1)$
values of the parameters $g$, $q$, and the hypercharges, for a
multiplicity of the gauge fields in the ballpark $N \sim
\mathcal{O}(10)$.

Although the aforementioned setup can be generally invoked for a wide
range of DM masses (\emph{i.e.} around weak-scale and beyond), we
focus in this work on a specific case of how the clockwork mechanism
can help realize a weak or near-weak scale CHAMP\footnote{Note that this is markedly different from an interesting DM scenario discussed in ref.\cite{Lee:2017fin} where the zero mode of the gauged clockwork sector is a \emph{dark} photon. In that case, it is the SM fermions that are millicharged under the residual dark $U(1)'$ symmetry and the DM remains electrically neutral (\emph{i.e.} with no couplings with the SM photon).}, motivated by the following attributes. Firstly, as argued before, this assumes
minimalism and thereby avoids large hierarchies with respect to the EW
scale. Secondly, a weak-scale CHAMP DM attracts the strongest
direct-detection limit on the fractional charge, namely $\epsilon
\lesssim 10^{-12}$, which makes it the most interesting minicharged DM
candidate to investigate against the role of a clockwork portal in
naturally realizing tiny charges. Finally, such a scenario may be
testable at some of the upcoming MeV scale gamma-ray telescopes (see \emph{e.g.} ref.\cite{Cirelli:2025qxx} for a list of relevant experiments and the corresponding projections).

The paper is structured as follows. Section~(\ref{sec:mod}) describes
the model in detail. In section~(\ref{sec:cons}), we discuss the various
theoretical and experimental constraints relevant for the
model. Following that, we illustrate the viability of a weak-scale
CHAMP as a DM candidate in section~(\ref{sec:DMpheno}) for a few
representative benchmark configurations. We finally summarize and
conclude in section~(\ref{sec:disc}).

\section{Model}
\label{sec:mod}
As described above, our goal is to dynamically generate a tiny
electrical charge for the DM particle. Since  $SU(2)$
charges are necessarily quantized, such a
  small charge can come about only by generating a minuscule
hypercharge for a $SU(2)$-singlet DM particle. To
this end, we begin by constructing a clockwork sector that would serve
to extend the hypercharge symmetry of the SM. The next step would be
to connect the standard electroweak sector and the clockwork sector,
taking care that the thus enhanced model is consistent with all
low-energy constraints. Finally, we bring in the dark matter
candidate, thereby completing the model. In the following, we
delineate each component individually.

\subsection{Clockwork with vector fields}
\label{ss:clvec}
Following the treatment in ref.\cite{Giudice:2016yja}, we define the
clockwork sector of the model with a gauge symmetry $U(1)^{N+1}$ ---
each factor corresponding to a gauge field $X_\mu^j$ which is
spontaneously broken, at a characteristic scale\footnote{In principle,
unequal SSB scales $f_i$ could be accommodated, but only at the cost
of introducing additional algebraic complexity without altering the
essentials. Large hierarchies between the $f_i$ would, however, negate
the spirit of the clockwork paradigm. To avoid such mundane
complications, we choose $f_i=f$, $\forall i$.}  $f$, to a single
$U(1)$ by a configuration of $N$ complex scalars $\Phi_j$. Responsible
for the \emph{clockworking} among the gauge fields, this configuration
is defined so as to have each scalar $\Phi_j$ charged under two
adjacent gauge groups, \emph{viz.} $U(1)_j \times U(1)_{j+1}$, with
the quantum numbers $(1,-q)$, where $q \gapp 1$.  The pertinent
Lagrangian can be expressed as\footnote{With fields carrying
  nonzero charges under more than one $U(1)$ groups, kinetic mixings
  between the corresponding gauge bosons are expected to be generated
  through quantum corrections. In the present case, we only have a
  single scalar shared between two adjacent groups.  Working in the
  unitary gauge, it is easy to see that the only relevant vertex is
  the four-point (two scalars and two gauge bosons) one, and the
  consequent diagrams do not contribute to the gauge-field
  wave-function renormalization and, hence, to kinetic mixing. The
  corrections to the masses of the heavy gauge bosons and their
  mixings are too small to be of any phenomenological consequence.}

\begin{equation} \label{eq:lcw}
  \mathcal{L}_{CW} = \ -\sum_{j=0}^{N} \frac{1}{4} X^j_{\mu\nu}  X^{j\, \mu\nu}
  + \sum_{j=0}^{N-1} \left[
(D_\mu \Phi_j)^\dagger (D_\mu \Phi_j) - \xi \left(\Phi_j^\dagger \Phi_j -\frac{f^2}{2}\right)^2\right] \, ,
\end{equation}
where $ X_{\mu \nu}^j$ denote the gauge field strengths and $\xi$ is a
dimensionless positive coupling. The covariant derivatives of the scalars are given by
\begin{align*}
    D_\mu \Phi_j &\equiv \left[ \partial_\mu +ig_x \left( X_\mu^j -q X_\mu^{j+1} \right)
      \right] \Phi_j \, ,
\end{align*}
where, for the sake of simplicity, we have assumed a
universal\footnote{Eschewing this assumption does not alter the
symmetry properties of theory. It will, in general, affect the masses
of the gauge fields and their effective couplings with the SM fermions
and the dark matter candidate, as will be evident shortly. However, in
the absence of large \emph{ad hoc} hierarchies in the gauge couplings,
the qualitative features of the model's phenomenology are adequately
addressed with the uniform parametrization.} gauge coupling $g_x$
across all the $U(1)$s.  Note that the structure of the Lagrangian is
reminiscent of moose or quiver
theories~\cite{Georgi:1985hf,Nelson:1985tx,Rothstein:2001tu,Arkani-Hamed:2002iiv,Gregoire:2002ra},
where the individual $U(1)$'s can be regarded as \emph{sites} and the
complex scalars as the conjoining \emph{link} fields.

The scalar potential in Eq.~\eqref{eq:lcw} implies that each of the $N$
$\Phi_j$'s receives a nonzero vacuum expectation value (VEV) given by
\beq 
\langle \Phi_j^{\dagger} \Phi_j \rangle = \frac{f^2}{2} \, 
\eeq
which spontaneously breaks the full gauge symmetry in the pattern
$U(1)^{N+1} \to U(1)_{CW}$,where the conserved charge is given
by\footnote{Had we considered nonuniform gauge couplings $g_i$, the
expression for $Q_{CW}$ would have changed, picking up factors of
$g_i$, but the core essence of the mechanism would have remained
unaltered.}
\begin{equation}
    Q_{CW} = \sum_{j = 0}^N q^{-j} Q_j \ ,
\end{equation}
with $Q_j$ denoting the individual $U(1)$ generators.
Consequently, a total of $N$ gauge fields
are Higgsed and receive masses while one combination remains massless,
corresponding to the unbroken symmetry $U(1)_{CW}$. The resulting
spectrum of the gauge fields is then easily obtained by considering the gauge Lagrangian in the unitary gauge,
\emph{i.e.}

\begin{equation}
  \mathcal{L}^{(2)}_{gauge}
  = -\sum_{j=0}^{N} \frac{1}{4}
   X^j_{\mu\nu}  X^{j\, \mu\nu}+ \sum_{j=0}^{N} 
   \frac{1}{2} \tilde M^2_{jk} X^j_\mu X^{k \mu}\ ,
\end{equation}
where the mass matrix is given by

\begin{equation}
\tilde{M}^2 = g_x^2 f^2 
    \begin{pmatrix}
1 & -q & 0 & \cdots &  & 0 \cr
-q & 1+q^2 & -q & \cdots &  & 0 \cr
0 & -q & 1+q^2 & \cdots & & 0 \cr
\vdots & \vdots & \vdots & \ddots & &\vdots \cr
 & & & & 1+q^2 & -q \cr
 0 & 0 & 0 &\cdots & -q & q^2
\end{pmatrix} \, .
\end{equation}
Diagonalizing $\Tilde{M}^2$ yields the
physical spectrum, namely

\begin{equation} \label{eq:cwev}
     m_0 = 0 \qquad {\rm and} \qquad m_k^2\equiv g_x^2 \ f^2 \lambda_k =
     g_x^2 \ f^2 \ \left(1+q^2-2 \ q \cos{\frac{k \pi}{N+1}} \right)
     \quad (k = 1\dots N)\, .
\end{equation}
There is, thus, a mass gap of ${\cal O}(g_x f)$ with the heavier
states being bunched relatively close\footnote{Note that, for
  large $q$, the mass gap is closer to $\sim g_x f |q - 1|$, a fact
  that would allow for a relatively smaller $f$ to be
  phenomenologically consistent.}. With $q$ being a real number, the
transformation between the gauge ($X_\mu$) and the mass basis
($B^0_\mu$) is simply realized by an orthogonal transformation,
\begin{equation}
    X^{\mu}_j = \sum_{k=0}^{N}O_{jk} B^\mu_k
\end{equation}
where the matrix $O$ is
specified as
\begin{equation}
\label{eq:mix}
    O_{j 0} = \frac{{\cal N}_0}{q^j} \, , ~~~ 
    O_{j k} = {\cal N}_k \left[ q \sin \frac{j k\pi}{N\! +\! 1}-  \sin \frac{  (j +1) k\pi}{N\! +\! 1} \right] \, ,~~~ j =0, .. , N;~~k =1, .. , N
\end{equation}
with
\begin{equation*}
    {\cal N}_0 \equiv \sqrt{\frac{q^2-1}{q^2-q^{-2N}}} \, , ~~~~ {\cal N}_k \equiv \sqrt{\frac{2}{(N\! +\! 1)\lambda_k}} \, .
\end{equation*}
For $q>1$,the massless state, $B^0_\mu$, is clearly localized towards
the $0$-th site, \emph{i.e.} has the maximum ($\sim \mathcal{O}(1)$)
overlap with $X_0^\mu$. Moreover, for sufficiently large $N$, it has
only an exponentially suppressed overlap of magnitude $\sim q^{-N}$
with $X^\mu_N$. Thus, a field external to the clockwork sector and
charged only under $U(1)_N$ will have but a tiny coupling proportional
to $q^{-N}$ with the zero mode $B^\mu_0$, effectively constituting a
small charge under $U(1)_{CW}$. We will, subsequently, make use of
this feature to naturally generate a very tiny electric charge for the
dark matter candidate while assuming all the coupling parameters of
the model to be at the $\mathcal{O}(1)$ level\footnote{While the
possibility of deriving tiny electromagnetic charges through the
clockwork mechanism was hinted at in ref.\cite{Giudice:2016yja}, our
model is the first explicit realization of the idea.}.

\subsection{The clockwork extended SM}

From the preceding discussion, it is clear that a small electric
charge for a $SU(2)$-singlet field localized at the $N$-th site of the
clockwork lattice can be naturally generated if the 
(conserved) CW symmetry could be identified with the hypercharge
component of the SM, {\em viz.} $U(1)_{CW} \equiv U(1)_Y$.  The small
overlap of the $B^0_\mu$ with the $X^N_\mu$ would then translate to an
effectively tiny hypercharge quantum number for such a field (without
the need to introduce a hierarchically small quantum number). In
doing this, it must be ensured, though, that the construction does not
run afoul of low-energy phenomenology, in particular the electroweak
precision tests. This is carried out systematically as follows.

\begin{table}
\begin{tabular}{||c c c c c c c c c c ||} 
 \hline
 Particles & $SU(2)_l$ & $SU(3)_c$ & $U(1)_0$ & $U(1)_1$ & ... & $U(1)_j$ & $U(1)_{j+1}$ & ... & $U(1)_N$ \\ [0.5ex] 
 \hline\hline
 $\psi_L$ & \textbf{2} & \textbf{1} & 1/2 & 0 & ... & 0 & 0 & ... & 0 \\ 
 $l_r$ & \textbf{1} & \textbf{1} & -1 & 0 & ... & 0 & 0 & ... & 0 \\ 
 $Q_l$ & \textbf{2} & \textbf{3} & 1/6 & 0 & ... & 0 & 0 & ... & 0 \\
 $u_r$ & \textbf{1} & \textbf{3} & 2/3 & 0 & ... & 0 & 0 & ... & 0 \\
 $d_r$ & \textbf{1} & \textbf{3} & -1/3 & 0 & ... & 0 & 0 & ... & 0 \\
 $H$ & \textbf{2} & \textbf{1} & 1/2 & 0 & ... & 0 & 0 & ... & 0 \\
 $\chi$ & \textbf{1} & \textbf{1} & 0 & 0 & ... & 0 & 0 & ... & $Y_\chi$ \\
 $\Phi_j$ & \textbf{1} & \textbf{1} & 0 & 0 & ... & 1 & $-q$ & ... & 0 \\ [1ex]
 \hline
\end{tabular}
\caption{Charge assignment under extended EW $+$ CW sector scenario.}
\label{tab1}
\end{table}

We start by assuming the full (\textbf{EW + CW}) symmetry of the model
Lagrangian to be $SU(2)_L \times U(1)^{N+1}$ with the corresponding
charge assignments of the SM fields as listed\footnote{As can be seen,
we have chosen a flavor-diagonal charge assignment. Consequently, no
tree-level flavor-changing neutral currents (FCNCs) are induced. Any
residual FCNC effects, if present, can arise only at higher orders and
are expected to be highly suppressed.} in Table (\ref{tab1}). Thus, in
addition to the usual $SU(2)_L$ charges, the SM matter fields
(including the Higgs field) are charged under $U(1)_0$, or in other
words, are localized at the $0$-th site of the CW lattice. Note that
the $U(1)_0$ charges for the SM fields are exactly the same as the
$U(1)_Y$ charges in the SM. To specify the model completely, we also
introduce at this point an extra Dirac fermion $\chi$ ( as the CHAMP
DM candidate) localized at the $N$-th site. The dynamics and viability
of $\chi$ as the DM candidate will be thoroughly discussed in a later
section. With these definitions, the electroweak Lagrangian in the
unbroken phase can be written as
\begin{equation}
  \label{eq:lag}
 \begin{split}
 \mathcal{L} &= \mathcal{L}_{Vis.} + \mathcal{L}_{CW} + \mathcal{L}_{H\Phi} + \mathcal{L}_{\chi} \, ,
 \end{split}   
\end{equation} 
where $\mathcal{L}_{Vis.}$ defines the gauge-invariant Lagrangian for
the SM fermions and the Higgs field $H$. On the other hand, the
component $\mathcal{L}_{H\Phi}$ specifies the mixing between the Higgs
and the CW scalars, and $\mathcal{L}_{\chi}$ defines the dark sector viz.,
\begin{equation}
  \label{eq:lag2}
 \mathcal{L}_{H\Phi} \equiv - \eta \sum_{j=0}^{N-1} H^{\dagger}H \Phi^{\dagger}_j\Phi_j \, \qquad {\rm and} \qquad 
 \mathcal{L}_{\chi} \equiv  \bar{\chi} \left(i\slashed{\mathcal{D}} - m_{\chi}\right) \chi
 \; .
\end{equation}
In the preceding, $\eta$ is a real dimensionless
coupling\footnote{We could, of course, have introduced
  nonuniform couplings $\eta_j$. This, however, would have only
  resulted in algebraic complexity without affecting the essential
  features of the scenario.}, $m_{\chi}$ denotes the mass of the
CHAMP and the covariant derivative $\mathcal{D}$ is given by
\beq  \label{eq:covchi}
\mathcal{D}_\mu \chi = \left(\partial_\mu + i g_x Y_\chi   X_\mu^N \right) \chi \, ,
\eeq
where the charge $Y_\chi$ is presumably ${\cal O}(1)$.

\begin{figure}
    \centering
    \includegraphics[width=0.8\linewidth]{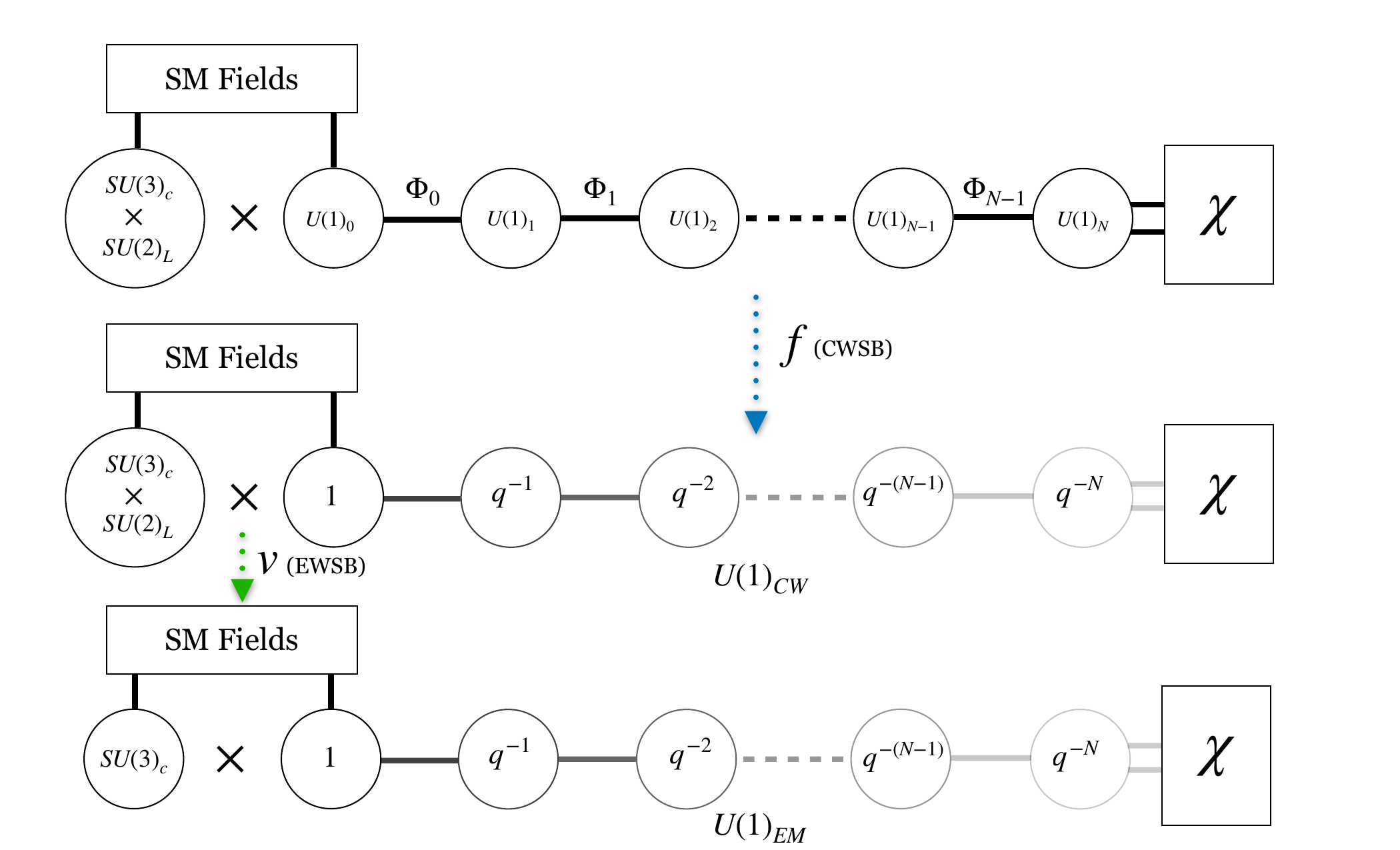}
    \caption{A schematic representation of the two-step symmetry
      breaking pattern in the model. The first row represents the
      extended hypercharge construction. In the second row, the
      individual blobs indicate the exponentially falling
      (site-dependent) charge under $U(1)_{CW}$. Similarly, in the
      third row, this translates into the strength of the photon
      couplings following EWSB (\emph{sans} the overall gauge
      couplings).}
    \label{fig:schematic}
\end{figure}

On both the CW scalars and the Higgs field
acquiring nonzero VEVs (at scales $f$ and $v$ respectively),
the symmetry of the theory spontaneously breaks down to a single
Abelian factor which is to be identified with the electromagnetic
symmetry, \emph{i.e.} $SU(2)_L \times U(1)^{N+1} \to U(1)_{EM}$. Of
course, the minimum of the full $(H,\Phi)$ potential depends on the
configuration of its parameters, the details of which we defer to a
later section. For now, we concentrate only on the gauge sector.
 The tree-level mass of the $W^\pm$ boson is
unaffected by the Abelian sector and, hence, can be ignored in the
discussion to follow. On the other hand, in the broken phase, the terms quadratic in the fields
$W^{3}_{\mu}$ and $X^{\mu}_j$  define a mass matrix given by
\begin{equation}
\label{eq:mass}
\mathcal{M}_{V}^2 = g_x^2 \  f^2
	\begin{pmatrix}
	 \frac{\delta}{4} b^2 & \frac{\delta}{4} b & 0 & 0 & 0 & ... & ... & 0 & 0 \\
	\frac{\delta}{4} b  & 1+ \frac{\delta}{4} & -q &  0 &  0 & ... & ... &  0 & 0\\
	0  & -q & 1+q^2 & -q & 0 & ... & ... & 0 & 0\\
	0  & 0 & -q & 1+q^2 & -q & \ddots & \ddots & \vdots & \vdots\\
	\vdots & \vdots & \vdots & \vdots & \vdots & \ddots & \ddots & \vdots & \vdots\\
    0  & 0 & 0 & 0 & ... & ... & 1+q^2 & -q & 0\\
    0  & 0 & 0 & 0 & ... & ... & -q & 1+q^2 & -q\\
    0  & 0 & 0 & 0 & ... & ... & 0 & -q & q^2\\ 
	\end{pmatrix}_{(N+2) \times (N+2)}
 \end{equation}
with
 \begin{equation} \nonumber
    \delta \equiv \frac{v^2}{f^2} \ , \quad \ b \equiv -\frac{g_w}{g_x} \, \, .
\end{equation}
  Here, $g_w$ is the $SU(2)_L$ gauge coupling.  It can be ascertained easily
   that ${\cal M}_V^2$ has one zero eigenvalue. As for the other eigenvalues, closed-form analytical expressions are not straightforward for the generic case. However, the task becomes reasonably simple for $v \ll f$ in which case the terms in
 $\mathcal{M}_V^{2}$ proportional to $\delta$ can be
 regarded as perturbations around the CW matrix $\tilde{M}^2$. In fact, from a symmetry-breaking perspective, the condition
 $v \ll f$ can be naively understood as a decoupling of the two SSB
 scales and, hence, the entire dynamics can be interpreted as a
 two-step process (see Fig.~(\ref{fig:schematic}) for an illustration), namely,
 \beq
\label{eq:brk_arr} 
  SU(2)_L \times U(1)^{N+1} \xrightarrow[]{f} SU(2)_L \times U(1)_{CW} \xrightarrow[]{v} U(1)_{EM} \, .
\eeq

After the first SSB occurs at the scale $f$, the light gauge boson
spectrum in the theory consists of the zeroth CW mode $B_{\mu}^{0}$
and the weak bosons $W_{\mu}^a$, all exactly massless. At this stage,
the field $B_{\mu}^{0}$ mimics the role of the SM $U(1)_Y$
boson. Understandably, then, one retrieves the usual dynamics of the
EW sector after the second SSB at the scale $v$, albeit with small
modulations induced by the CW extension. That there indeed exists a
region in the parameter space where deviations as such fall within the
known uncertainties in the couplings and masses of the EW gauge bosons
will be shown in the next section. For now, we delineate the details
of the mass eigenvalues.


Post EWSB, the quadratic terms involving the full set of neutral gauge
bosons (in the $W^3 - B^{k}$ basis) are given by

\beq
\mathcal{L}^{(2)}_{neut.} \supset \frac{v^2}{8} \left(-g W^3_{\mu} + g_x \sum_{j=0}^{N} O_{0 j}B^{j}_{\mu} \right)^2 \, + \, \mbox{(CW mass terms)} \, .
\eeq
where the matrix elements $O_{0 j}$ are as given in
Eq.~\eqref{eq:mix}.   It is convenient to first identify the
photon ($A_{\mu}$) through a rotation in the \((W^3,\,
B^{0})\) plane which reads 
\beq \label{eq:gamma}
A_{\mu}
\equiv
\sin\theta_W \, W^3_\mu
+
\cos\theta_W B^{0}_{\mu} \, ,
\qquad {\rm and} \qquad
\tilde Z_{\mu}
\equiv
\cos\theta_W \, W^3_\mu
- \sin\theta_W B^{0}_{\mu} \, ,
\qquad
\eeq
The angle $\theta_W$, quite clearly, acts as an analogue of the
Weinberg angle in the SM, and is defined here as \( \tan \theta_W = (
g_xO_{00}/g_w)\). 
In terms of the rotated fields, the mass-squared
matrix can be written as
\begin{equation} \label{eq:mass2}
\tilde{\mathcal{M}}_V^2
\;\equiv\;
U^{T}\, \mathcal{M}_V^2\, U
\end{equation}
where the transformation matrix $U$ is defined as
\beq\label{eq:trans2}
    \begin{pmatrix}
        W^3_\mu\\ X^0_\mu\\ X^1_\mu\\ \vdots\\ X^N_\mu
    \end{pmatrix}
    =
    \begin{pmatrix}
        \cos\theta_W & \sin\theta_W & 0 & \cdots & 0 \\
        -\sin\theta_W O_{00} & \cos\theta_W O_{00} & O_{01} & \cdots & O_{0N} \\
        -\sin\theta_W O_{10} & \cos\theta_W O_{10} & O_{11} & \cdots & O_{1N} \\
        \vdots & \vdots & \vdots & \ddots & \vdots \\
        -\sin\theta_W O_{N0} & \cos\theta_W O_{N0} & O_{N1} & \cdots & O_{NN}
    \end{pmatrix}
    \begin{pmatrix}
       \tilde Z_\mu \\ A_\mu\\ B^1_\mu\\ \vdots\\ B^N_\mu
    \end{pmatrix} \, .
 \eeq
Given that $A_\mu$ is strictly massless, the (nondiagonal) matrix
$\tilde{\mathcal{M}_V^2}$ has rank $(N+1)$ and can be trivially
reduced to
\begin{equation} \label{eq:mass3}
\tilde{\mathcal{M}}_V^2
\; \to \;
\begin{bmatrix}
m_{\tilde Z}^2 & \Delta^{T} \\
\Delta  & \mathcal{M}'^{2}
\end{bmatrix}_{(N+1)\times (N+1)} \ .
\end{equation}
Here,
\begin{equation} \label{eq:z0mass}
m_{\tilde Z}^2 \equiv \frac{v^2}{4}\left(g_x^2 O_{00}^2 + g_w^2\right) \, ,  
\end{equation}
alongwith the aforementioned identification $g_Y = g_{x}O_{00}$, would have
  denoted the tree-level mass of the SM $Z$-boson in the absence of
  mixings with the higher $B_k$'s ($k, l = 1\dots N$) as encoded in
\begin{equation} \label{eq:zzpmixings}
    \begin{split}
        \Delta_k &= -\frac{g_x g_w}{4\cos\theta_W}\,v^2\,O_{0k} \, , \\
        \left(\mathcal{M}'^2\right)_{kl} &= g_x^2 f^2 \lambda_k\,\delta_{kl}
        + \frac{g_x^2 v^2}{4}\,O_{0k}O_{0l} \, .
    \end{split}
\end{equation}
To leading order in $v^2/f^2$, the $Z$ boson is, then, identified as
\begin{equation} \label{eq:zmixeq}
    Z_\mu \simeq  \tilde Z_\mu
    + \frac{\sqrt{2} \, v^2 O_{00}}{4 f^2 \sqrt{N+1} \sin\theta_W}
    \sum_{k=1}^{N}
    \left( \frac{1}{ \lambda_k^{3/2}}\,\sin\frac{k\pi}{N+1} \right)
    B_\mu^k + \mathcal{O}\!\left(\frac{v^4}{f^4}\right)
\end{equation}
and the corresponding mass eigenvalue given by
\begin{equation} \label{eq:Zmass}
    m_Z \simeq  m_{\tilde Z} \left[
    1 - \frac{v^2}{4 f^2 (N+1)}
    \sum_{k=1}^{N}
    \frac{1}{\lambda_k^{2}}
    \sin^{2}\frac{k\pi}{N+1}
    \right]
    + \mathcal{O}\!\left(\frac{v^4}{f^4}\right) \, .
\end{equation}

As can be seen from Eq.~\eqref{eq:Zmass}, the deviation depends not
only on the hierarchy between EWSB and CWSB scales, parametrized by
\(v^2/f^2\), but also on the factor \(q\) through \(\lambda_k\). This dependence in nontrivial---see eqn.(\ref{eq:cwev})---and, for large $q$ (as we would discuss in the next section, $q \gtrsim 3$ is phenomenologically preferred) the deviation scales approximately as \(\sim v^2/(q^4 f^2)\).

While the obtained deviations from the SM expectations are crucial for
the $Z$ boson, as discussed in the next section, the corresponding
corrections for the heavier gauge bosons can safely be neglected, and
the corresponding physical states along with their masses can be well
approximated as
\begin{equation} \label{eq:heavyz}
    Z'^k_\mu \simeq B_\mu^k, 
    \qquad
    m_k \simeq g_x f \sqrt{\lambda_k} \, .
\end{equation}
With the above identification, the couplings of the neutral gauge
bosons to the SM currents $J_{\text{SM}}^f$ (for a given fermion $f$), as well as to the dark matter particle $\chi$, can be parametrized in terms of vector and axial-vector couplings $g^{f,i}_{V/A}$
\begin{align} \label{eq:smcouple}
\begin{split}
    g^{f,i}_{V/A} &=  \frac{1}{2}\left(g_x~ U_{1i}~  (~Y_L^f \pm Y_R^f) +  g_w ~ U_{0i} ~T^{3,f}\right)\\
    g^{\chi,i} &=  g_x~ U_{Ni} ~Y_\chi \ ,
\end{split}
\end{align}
where we have omitted the higher order corrections in $v^2/f^2$.

Contrary to the usual convention of labeling a multi-field spectrum according to the hierarchy in masses, we choose to label the lightest modes, \emph{i.e.} the photon and the $Z$ boson, according to their overlaps with the \emph{pure} SM vector fields in their gauge basis---in our context this happens to be the $W^3$ field.  With this prescription, therefore, the zeroth mode corresponds to the $Z$ boson while the first mode corresponds to the photon, and the heavier $Z'_k$ modes are then labeled according to their masses.
Thus, for any fermion \(f\), Eq.~\eqref{eq:smcouple} gives
\begin{equation} \label{eq:smemcoup}
    g_{V}^{f,1} = e Q_f \quad; \quad g_A^{f,1}=0
\end{equation}
where $Q_f = Y_L^f + T^{3,f} = Y_R^f$ denotes the electromagnetic
charge of the fermion $f$. Similarly, the $\gamma W^+ W^-$ coupling
too remains unaltered, while for DM, the electromagnetic coupling is given by
\begin{equation} \label{eq:dmemcoup}
    g^{\chi,1} =Y_\chi  \,  q^{-N} \,  e \equiv \epsilon \, e  \ .
\end{equation}
For the $Z$ boson (corresponding to the index
``0'' in Eq.~\eqref{eq:smcouple}), the tree-level vector and
axial-vector couplings are given by
\begin{equation} \label{eq:ztreecou}
    \begin{split}
        g_V^{f,0} &= \frac{g_w}{\cos\theta_W}
        \left[ \left(T^{3,f} - 2 Q_f \sin^2\theta_W \right) + \delta g_V^{f,0} \right] , \\
        g_A^{f,0} &= \frac{g_w}{\cos\theta_W}
        \left[ T^{3,f} + \delta g_A^{f,0} \right] .
    \end{split}
\end{equation}
The first terms on the right-hand side of Eq.~\eqref{eq:ztreecou}
correspond to the leading-order SM couplings obtained from
Eq.~\eqref{eq:smcouple}, while $\delta g_{V/A}^{f,0}$ denote the
corrections induced by the CW sector, and are given by
\begin{equation} \label{eq:vargz}
    \delta g_{V/A}^{f,0}
    = - \frac{v^2}{4 f^2} \,
    \frac{(Y_L^f \pm Y_R^f)}{(N+1)}
    \sum_{k=1}^{N}
    \frac{1}{\lambda_k^{2}}
    \sin^{2}\frac{k\pi}{N+1} \, .
\end{equation}

Similar to the mass, the deviation in the coupling also scales as \(\sim v^2/(q^4 f^2)\) for sufficiently large \(q\). As before, we neglect variations in the effective couplings of the
heavier CW gauge bosons to the SM sector, since these effects are
strongly suppressed and do not lead to qualitative changes in the
phenomenology.

\section{Constraints}
\label{sec:cons}

While our model does have many fields beyond the SM, it is quite
economic in terms of the new parameters, having only seven additional
ones. These include the clockwork factor \( q \), which controls the
exponential localization of modes across the clockwork sites and,
thereby, determines the apparently hierarchical structure of
couplings; the number of vector clockwork sites \( N \), which governs
the resolution and extent of the clockwork chain; and the clockwork
symmetry breaking (CWSB) scale \( f \), which sets the mass scale of
the heavy states arising from the clockwork sector.

Additionally, the quartic coupling \( \xi \) describes
self-interactions within the heavy scalar sector, whereas the coupling
\( \eta \) dictates the interaction strength between the Standard
Model Higgs and the clockwork scalars, leading to mixing
  between them\footnote{Note that the gauge coupling $g_x$ is not a
new parameter, having replaced the usual hypercharge coupling in the
SM.}. Finally, the dark matter sector itself is characterised by two
parameters, the mass \(m_\chi \) and \( Y_\chi \), its charge under
the extremal $U(1)$.

As with any model going beyond the SM, the new fields and parameters
are subject to both experimental as well as theoretical constraints
and it is useful to discuss these individually.

\subsection{Theoretical constraints}\label{subsec:theocons}
The theoretical constraints, as can be imagined, arise primarily in
the context of the high-energy behaviour of the multi-scalar potential
which, in turn, crucially dictates the validity of the clockwork
mechanism in the far UV. In the following, therefore, we delineate the
various elements of the scalar sector in the model and assess the
stability of the configuration under renormalization
  group (RG) evolution.

The full scalar potential of the theory reads
\begin{equation} \label{eq:pot}
  V(H,\{\Phi_j\}) = \lambda \left(H^\dagger H- \frac{v^2}{2}\right)^2
  + \xi \  \sum_{j=0}^{N-1} \left( \Phi_j^\dagger \Phi_j
  -\frac{f^2}{2}\right)^2
  + \eta  \ H^\dagger H  \sum_{j=0}^{N-1} \Phi_j^\dagger \Phi_j \ ,
\end{equation}
and for it to be bounded from below, one needs 

\beq \label{eq:potbound}
\lambda, \xi  > 0 \qquad \mbox{and} \qquad N\eta^2 < 4 \lambda \, \xi \, .
\eeq
The last-mentioned is an important
constraint as would, shortly, be evinced below.

In the unitary gauge, the fields may be expressed in terms
of the vacuum expectation values \(v\) and \(f\) as
\begin{equation}
    H = \begin{pmatrix} 0\\ \frac{\tilde{h}+v}{\sqrt{2}} \end{pmatrix}, \quad
    \Phi_j = \frac{\tilde{\phi}_j+f}{\sqrt{2}} \, .
\end{equation}
Clearly, the field \(\tilde{h}\) mixes only with one specific combination  of
the $\tilde \phi_j$, namely
\begin{equation} \label{eq:lincomb}
\tilde{\Phi} = \frac{1}{\sqrt{N}} (\tilde{\phi}_0 + \tilde{\phi}_1 + \dots + \tilde{\phi}_{N-1}).
\end{equation}
Consequently, the matrix is trivially diagonalized with the eigenvalues given by
\beq
\begin{split}\label{eq:eigvheavyscal}
m_h^2 &= \lambda v^2 + \xi f^2 - \sqrt{\left(\xi f^2 - \lambda v^2\right)^2 + \frac{N}{4} \eta^2 f^2 v^2},  \\ 
m_{\Phi}^2 &= \lambda v^2 + \xi f^2 +  \sqrt{\left(\xi f^2 - \lambda v^2 \right)^2 + \frac{N}{4} \eta^2 f^2 v^2}, \\
m_{\varphi_{k}}^2 &=  2 \xi f^2 \qquad (k=1,..,N-1) \, . \\
\end{split} 
\eeq
where $\varphi_k$ are orthogonal to $\tilde \Phi$ and to each other. A
simple identification would be with the diagonal generators of
$SU(N)$, but with these states being
degenerate, any linear combination is as good. As the expressions for
\(\varphi_i\)s are not relevant for our analysis, we shall consider
these no further.  The remaining two mass eigenstates $h$ and $\Phi$
are easily identified to be

\begin{equation}
    \begin{pmatrix}
        h \\ \Phi
    \end{pmatrix}
    =
    \begin{pmatrix}
        \cos \zeta & -\sin \zeta \\ \sin \zeta & \cos \zeta
    \end{pmatrix}
    \begin{pmatrix}
        \tilde{h} \\ \tilde{\Phi}
    \end{pmatrix}
\end{equation}
with the mixing angle \(\zeta\) being given by
\begin{equation} \label{eq:hsmixtan}
\tan 2\zeta = \frac{\sqrt{N} \, \eta f v}{2\left(\xi f^2 - \lambda v^2\right)} \, .
\end{equation}
In the limit of $v \to 0$, not only does the mixing disappear,
  but $\Phi$ becomes degenerate with the $\varphi_{k}$, as it
  should. This is only a reflection of the $SU(N)$ symmetry in the
  pure $\Phi_j$ sector, broken only by the other interactions.

Note that the aforementioned bound on $N \eta^2$
Eq.~\eqref{eq:potbound} not only serves to keep $m_h^2$ positive---see
the first of Eqs.\eqref{eq:eigvheavyscal}---but also controls the
Higgs mixing angle, which cannot be too large. In addition, the
stability condition must hold true at least upto a putative UV cutoff
scale as the parameters of the potential undergo RG
evolution. Understandably, this places the strongest theoretical upper
bound on $\eta$ in the IR (\emph{i.e.} around the scale $f$).

The aforementioned constraints have been imposed only on the
tree-level parameters. One must also ensure not only that the
couplings do not become nonperturbative under RG evolution but also
that the stability of the potential is not compromised until a cutoff
scale $\Lambda \gg f$. Limiting ourselves to only a one-loop analysis,
we have
\begin{align} \label{eq:betafunc}
  16 \pi^2 \beta_{\lambda}  \approx & \quad \frac{3}{8} g_x^4 + \frac{3}{4} g_x^2 g_w^2 + \frac{9}{8} g_w^4 - 3 g_x^2 \lambda - 9 g_w^2 \lambda + 24 \lambda^2 + N \eta^2 + 12 y_t^2-6 \ y_t^4 \nonumber\\
 16 \pi^2 \beta_{\xi} \approx & \quad 6 g_x^4(1+q^2)^2-12 g_x^2(1+q^2)\xi + 2 \eta^2
+ 20 \ \xi^2   \nonumber \\
16 \pi^2 \beta_{\eta} \approx & \quad 3 g_x^4 + 4 \eta^2
+\eta\left( - \frac{15}{2} g_x^2 - \frac{9}{2} g_w^2 + 12 \lambda 
+ 8 \xi - 6 g_x^2 q^2 + 6 y_t^2 \right)
\end{align}
Here $\beta_p \equiv \mathrm{d}p/\mathrm{d}t$, with
$t=\ln(\mu/\mu_0)$, where $\mu$ being the
renormalisation scale and $\mu_0 = m_t$ the reference scale at which
the boundary conditions are defined. While the evolutions of the nonabelian
  gauge couplings remain unaltered, that for the $U(1)$ factors is
  more nuanced. For $k = 1\dots (N-1)$, the gauge boson $X_k$ couples
  only to the scalars $\Phi_{k-1}$ and $\Phi_k$ carrying charges $-q$
  and $1$ respectively. With the scalars having equal masses, the
  corresponding gauge couplings evolve identically. The story for
  $U(1)_0$ and $U(1)_N$ is different. The first couples to all of the
  SM fermions, the Higgs field $H$ as well as the scalar
  $\Phi_0$. Hence, the evolution of this coupling is very similar to
  that for the $U(1)_Y$ inside the SM. The $U(1)_N$, on the other
  hand, couples to only $\Phi_{N-1}$ (with a charge $-q$) and the DM
  $\chi$ (with a charge $Y_\chi$), and so its coupling evolves
  differently from all others. This, of course, brings into question
  our simplifying assumption of all the $U(1)$ field having identical
  coupling constants. However, given that the evolutions are slow and
  the cutoff $\Lambda$ for the theory not too large (see the next
  subsection), if the couplings at the scale $\Lambda$ are held
  identical, $g_x(0)$ and $g_x(N)$ would differ from the rest only
  minimally. As we have argued earlier, the consequent changes in the
  phenomenology would be only quantitative, and that too minimal.

The evolution of $\lambda$ receives a positive contribution from the
Higgs-CW sector interaction through the $N\eta^{2}$ term. For large
$N$ or sizable coupling $\eta$, this contribution can enhance the
running of $\lambda$, thereby partially compensating for the
destabilising effect induced by the top Yukawa coupling. On the other
hand, too large a value for $N\eta^2$ would, nominally, tend to render
$\lambda$ nonperturbative. Fortunately, the value of $N\eta^2(t = 0)$
required to do this is too large to be of phenomenological interest.
Consequently, the overall behaviour of $\lambda$ depends sensitively
on the chosen benchmark parameters.

The evolution of the CW-sector self-coupling
  $\xi$ is dominantly driven by gauge interactions through the term
proportional to $g_x^{4}(1+q^{2})^{2}$, reflecting its strong
sensitivity to the parameter $q$. In addition, the positive
self-interaction contribution $20\xi^{2}$ leads to a rapid growth of
$\xi$ at high scales, potentially driving it towards a Landau pole for
sufficiently large initial values. The term proportional to $\eta^{2}$
remains subdominant for $\eta < \xi$, which is  natural
choice considering in view of the stability condition
of \eqref{eq:potbound}. 

Finally, the running of the Higgs-CW coupling $\eta$ is governed by a
balance between scalar and gauge effects. Positive contributions arise
from scalar self-interactions (involving the parameters $\lambda$,
$\xi$, and $\eta$), as well as from the top Yukawa coupling $y_t$,
whereas gauge interactions induce negative contributions. In
particular, the term proportional to $\xi$ becomes dominant and leads
to an accelerated enhancement of $\eta$ in the ultraviolet. For
moderate starting values of $N\eta^2$ and $\xi$, the change in $\eta$
is confined to a few percent for a cutoff $\Lambda \sim
10^3$~TeV. Given this, it is easy to see that the conditions of $\xi >
0$ is never violated and even $N\eta^2 < 4 \lambda \xi$ is not
violated until one almost hits the epoch of
metastability of the SM Higgs (which, as we have argued above, is
postponed compared to the case of the SM).

In summary, while the couplings $\eta$ and $\xi$ are simultaneously
constrained by Higgs--scalar mixing limits, vacuum stability, and
triviality requirements, the restrictions are not too severe. With
$\lambda \simeq 0.12$, $v \simeq 246~\mathrm{GeV}$ being given, and
even allowing for a maximal Higgs mixing with a singlet scalar {\em
  viz.} $\sin\zeta \lesssim 0.1$~\cite{Robens:2016xkb},
Eq.~\eqref{eq:hsmixtan} implies $\eta \lesssim 0.1$ for $\xi \simeq
0.7$, assuming $f \simeq 1~\mathrm{TeV}$ and $N=20$. This constraint
is further relaxed for larger values of $f$, while the RG evolution
governed by Eq.~\eqref{eq:betafunc} remains perturbative and safely
below the triviality bound up to scales of at least $10^3$~TeV.

\subsection{Experimental constraints}
Confronting the parameter space of our model with experimental data, the most stringent constraints emanate from sources like electroweak precision measurements, collider searches (especially in the dilepton channel), and direct detection experiments targeting dark matter. Electroweak observables, measured with high accuracy at colliders, are particularly sensitive to new physics that modifies the gauge boson sector as well as the scalar sector, while direct detection bounds constrain the allowed strength and structure of interactions between the DM and the SM particles.

\subsubsection{Electroweak Constraints}

First and foremost, note that $g_w$---the $SU(2)_L$ gauge
coupling---remains the same as in the SM; and while \(M_W\) may
receive small radiative corrections from the clockwork \(Z'\) bosons,
such corrections would turn out to be negligible due to the suppressed
coupling of \(W^3\) to the heavy clockwork states, and can be safely
ignored. Thus we may safely retain the SM relation
\begin{equation}
\label{eq:cons3}
    g_w^2 = \frac{8 G_F M_W^2}{\sqrt{2}}.
\end{equation}
Further, the identification of $g_Y$ elucidated in the preceding section, \emph{viz.}
\begin{equation}
\label{eq:cons1}
    g_x O_{00} = g_x \sqrt{\frac{q^2 - 1}{q^2 - q^{-2N}}} \equiv g_Y = \frac{e}{\cos \theta_W},
\end{equation}
implies a relation between the parameters \(g_x\), \(q\), and
\(N\). The relation above could also be rewritten as
\begin{equation}
\label{eq:cons2}
   e = \frac{g_x  g_w O_{00}}{\sqrt{g_x^2 O_{00}^2 + g_w^2}}.
\end{equation}

Reminding ourselves that the DM coupling to the photon---given by
$g^{\chi,1} = e q^{-N} Y_\chi$---has to be tiny, $q^N$ needs to be very
large unless $Y_\chi$ is unnaturally small. This, in turn, means that $N$
essentially decouples from the relation in Eq.~\eqref{eq:cons1} and
$O_{00} \simeq \sqrt{1 - q^{-2}}$. In other words, the fine structure
constant essentially imposes a relation between $g_x$ and $q$, almost
independent of the other parameters in the theory.

\begin{figure} 
    \centering
    \includegraphics[width=0.66\linewidth]{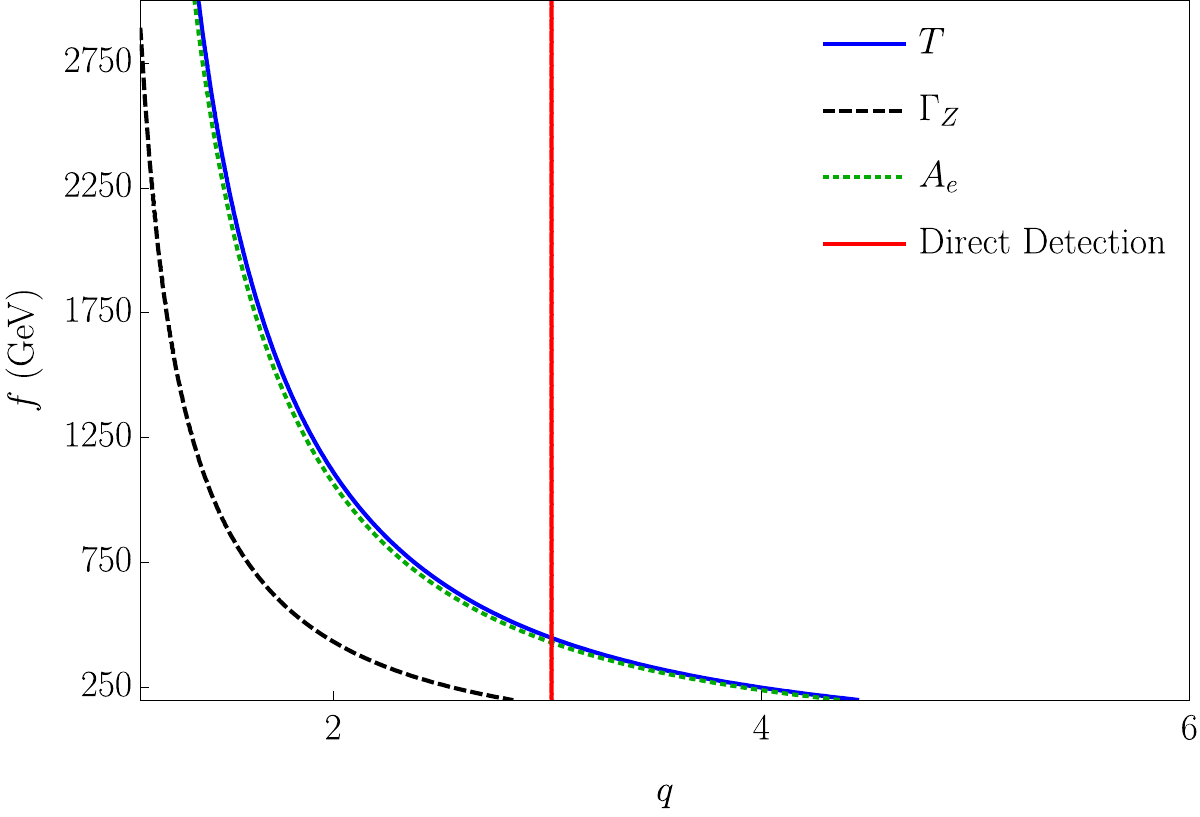}
    \caption{Bounds from $Z$-pole observables in the $(q,f)$ parameter space for fixed $N=25$ at the \(68 \%\) CL. The region to the above each curve is allowed. The black dashed curve corresponds to the constraint from the $Z$-boson decay width, the blue solid curve arises from the oblique parameter $T$ (we have taken \(T= 0.01 \pm 0.12\) \cite{ParticleDataGroup:2024cfk}), and the green dashed curve represents the constraint from the electron left-right asymmetry. The red vertical line indicates the lower bound on $q$ for the given $N$, as required by direct detection limits on fractionally charged DM (discussed in the next section).}
    \label{fig:zpoleobs}
\end{figure}

At this stage, we are in a position to examine the electroweak
  precision observables. As shown above, both the mass and the
  couplings of the \(Z\) boson receive tree-level modifications from
  the CW sector, as given in Eqs.~\eqref{eq:Zmass} and
  \eqref{eq:vargz}, respectively. We now move to the well-known electroweak precision observables, and
begin by discussing the shift in
  the \(Z\)-boson mass, which can be expressed in terms of the oblique
  parameter \(T\), defined as
\begin{equation} \label{eq:tparam}
    T = \frac{-1}{\alpha}\,\frac{\Pi^{\text{new}}_{ZZ}(0)}{m_Z^2} \, .
\end{equation}
Here, $\alpha$ is the electromagnetic coupling evaluated at the $Z$
pole, and $\Pi^{\text{new}}_{ZZ}(0) = m_Z^{2} - (m_{\tilde Z})^2$. Since no tree-level corrections accrue to the
$W$-boson mass, the $T$ parameter, in our setup, can be split neatly into the tree-level contribution from the correction to the
  $Z$-boson mass and the usual one-loop corrections within the SM. In this, we are neglecting the quantum
corrections accruing from the new physics particles in the
  loops, as these are too small to be of any consequence. We then have,
\begin{equation}\label{eq:Tparam}
    \alpha T_{\rm tree} = \frac{v^2}{2 f^2 (N+1)}  \sum_{k=1}^{N}  \frac{1}{\lambda_k^{2}}  \sin^{2}\frac{k\pi}{N+1} \, ,
\end{equation}
whereas the expression for $T_{\rm SM}$ remains
  unchanged. $T_{\rm tree}$ has the most pronounced dependence on the
  parameter $q$ and $f$, and, in Fig.~(\ref{fig:zpoleobs}), we display
  the resulting exclusion regions for a fixed $N$.

  Of equal importance are the $Z$-couplings to the SM
    fermions. Within the SM, the {\em observable} Weinberg angle
    $\tilde \theta_W$ is expressed in terms of the parameter
    appearing in the Lagrangian through~\cite{Appelquist:2002mw}
\begin{equation}
    \sin^2\tilde{\theta}_W \cos^2\tilde{\theta}_W
    = \sin^2\theta_W \cos^2\theta_W \left(1 + \alpha T_{\rm tree}\right) \, .
\end{equation}
The effective couplings corresponding to the tree-level
  ones--see Eq.~\eqref{eq:ztreecou}---can, then, be parametrized as
\begin{align} \label{eq:coupZeff}
    \tilde g_V^{f,0} &=
    \frac{e}{\sin\tilde{\theta}_W \cos\tilde{\theta}_W}
    \left( \tilde g^{f,0}_{V,\mathrm{SM}} + \delta \tilde g^{f,0}_V \right) , \\
    \tilde g_A^{f,0} &=
    \frac{e}{\sin\tilde{\theta}_W \cos\tilde{\theta}_W}
    \left( \tilde g^{f,0}_{A,\mathrm{SM}} + \delta \tilde g^{f,0}_A \right) .
\end{align}
where $\tilde g^{f,0}_{V,\mathrm{SM}} = T^{3,f} - 2 Q_f \sin^2\tilde{\theta}_W$ and
$\tilde g^{f,0}_{A,\mathrm{SM}} = T^{3,f}$ denote the SM contributions to the vector and axial-vector couplings, respectively. The deviations $\delta \tilde g^{f,0}_{V,A}$ encode the effects of new physics and are given as
\begin{align} \label{eq:varZeff}
    \delta \tilde{g}^{f,0}_{V} &= \frac{1}{2} \alpha T_{\rm tree} \left( \tilde g^{f,0}_{V,SM} + Q_f \ \frac{\sin \tilde \theta_W \cos \tilde \theta_W}{\cos^2 \tilde \theta_W - \sin^2 \tilde \theta_W} \right)+ \delta g_{V}^{f,0} \quad , \\
    \delta \tilde{g}^{f,0}_{A} &= \frac{1}{2} \alpha T_{\rm tree} \left( \tilde g^{f,0}_{A,SM}  + Q_f \ \frac{\sin \tilde \theta_W \cos \tilde \theta_W}{\cos^2 \tilde \theta_W - \sin^2 \tilde \theta_W} \right)+ \delta g_{A}^{f,0} \quad ,
\end{align}
with $\delta g^{f,0}_{V/A}$ denoting the tree-level deviations from the SM induced by the
new physics, as given in Eq.~\eqref{eq:vargz}.

While a global fit to all electroweak precision data is
  possible, it suffices to consider the measurements with the highest
  sensitivities, namely the total decay width of
the $Z$ boson, $\Gamma_Z$,
\begin{equation} \label{eq:gammaZexp}
    \Gamma_Z= \Gamma^{SM}_{Z} \left[1+ 2 \sum_f \frac{\tilde g^{f,0}_{V,SM} \  \delta \tilde{g}^{f,0}_{V}+ \tilde g^{f,0}_{A,SM} \ \delta \tilde{g}^{f,0}_{A} }{(\tilde g^{f,0}_{V,SM})^2 + (\tilde g^{f,0}_{A,SM})^2}\right] \, .
\end{equation}
and the left-right asymmetry $A_e$ as
  measured by the SLD and LEP,
\begin{equation} \label{eq:LRasymm}
    A_e= A_{e,SM} \left[1+  4 \frac{(\tilde g^{e,0}_{L,SM})^2 \  (\tilde g^{e,0}_{R,SM})^2  }{(\tilde g^{e,0}_{L,SM})^4 - (\tilde g^{e,0}_{A,SM})^4} \left(\frac{\delta \tilde{g}^{e,0}_{L}}{g^{e,0}_{L,SM}}-\frac{\delta \tilde{g}^{e,0}_{R}}{g^{e,0}_{R,SM}}\right)\right] \, .
\end{equation}
with the left- and right-handed couplings obtained from the vector and
axial couplings as \(\tilde g^{e,0}_{L/R} = \tilde g^{e,0}_{V} \pm
\tilde g^{e,0}_{A}\). The quantities \(\Gamma^{\mathrm{SM}}_{Z}\) and
\(A_{e,\mathrm{SM}}\) denote the Standard Model predictions for the
$Z$-boson decay width and the LR asymmetry, respectively, including
radiative corrections. The current experimental measurements,
\(\Gamma_Z^{\mathrm{exp}} = 2.4955 \pm 0.0023\) and
\(A_e^{\mathrm{exp}} {(\rm LEP)} = 0.1498 \pm 0.0049\)
\cite{ParticleDataGroup:2024cfk}, are in excellent agreement with the
SM results, thereby placing stringent constraints as depicted in
Fig.~(\ref{fig:zpoleobs}). As can be seen, the constraints imposed by
the individual EW observables ($T$, $\Gamma_Z$ and $A_e$) on the
$(q-f)$ plane are largely similar in effect, with the ones resulting
from $\Gamma_Z$ being slightly weaker. For
reference, we also indicate the direct detection bound on \(q\) for a
fixed \(N\), corresponding to the allowed region for millicharged dark
matter, with a red vertical line, which will be discussed in the next
section.

As for the parameters of the extended scalar sector, the experimental
measurements primarily constrain the $\Phi-h$ mixing angle
$\zeta$. These bounds arise from a number of sources, namely the
electroweak precision observables; the measured $W$ mass; the observed
signal strengths of the 125~GeV Higgs resonance at ATLAS and CMS; as
well as limits from SM Higgs invisible decays (for light new
scalars). Also relevant are constraints from direct searches for new
scalar resonances at the LEP, Tevatron, and the LHC. The
aforementioned constraints have been thoroughly investigated in the
literature
\cite{Pruna:2013bma,Robens:2015gla,Bojarski:2015kra,Robens:2016xkb,Robens:2025nev,Lewis:2024yvj,Lane:2024vur,Adhikari:2020vqo}
and are directly applicable to our setup.  For the new scalar masses
in the range $ 2 m_h \, \mbox{GeV} \lesssim m_{\Phi} \lesssim 600 \,
\mbox{GeV}$, the bounds on $\sin \zeta$ are dominated by those from
direct searches for $\Phi \to
ZZ$, $\Phi \to VV$ and $\Phi \to h h$, whereas for $m_{\Phi} \gtrsim
600$ GeV constraints from higher-order contributions to the $W$ mass
prevail. In both the cases, though, the upper limit amounts to $|\sin
\zeta| \lesssim 0.15$ \cite{Feuerstake:2024uxs, Robens:2025nev}.
    
Additionally, for the opposite case of $m_h < m_{\Phi} <  2m_h$, strong constraints on the
mixing angle apply from the
exclusion limits from Higgs searches at LHC \cite{Robens:2016xkb}, whereas for $m_{\Phi} < m_h$, the same is constrained by measurements of the Higgs signal strength at LEP \cite{ALEPH:2006bhb}. However, given the current
experimental lower limits on the $Z'$ masses in our model (to be
discussed later), the allowed CWSB scale can be roughly estimated to
be $f \gtrsim 1$ TeV. Consequently, with the assumption that $\xi \sim
\mathcal{O}(1)$, the heavy CW scalars are naturally expected to have
commensurate masses, \emph{i.e.} $m_{\Phi} \gtrsim 1$ TeV. With this
in mind, we do not explore the case $m_{\Phi} < 2 m_h$ any further.

\subsubsection{Direct detection constraints}
\label{ssec:ddc}

Direct detection experiments largely employ bolometric devices
dependent on energy transferred in the DM particle recoiling off the
detector atoms (essentially the nuclei). Consequently, non-detection places upper
bounds on the recoil cross sections.

In the present context, this is driven by a series of $t$-channel
diagrams, with the mediators being all the neutral gauge bosons in the
theory (the photon, the $Z$, and all of the $Z'$s). The amplitude, for
scattering off a quark $q$, can be represented as
\[
  {\cal M} = t ^{-1} J_{1 \mu}^{(\chi)} J_{1}^{(q)\mu} + (t - m_Z^2)^{-1} J_{0 \mu}^{(\chi)} J_{0}^{(q)\mu}+\sum_{k = 1}^{N} \left(t - m_k^2\right)^{-1} J_{k+1, \mu}^{(\chi)} J_{k+1}^{(q)\mu}
\]
where the appropriate gauge couplings have been subsumed in the
currents. In the above, due to the vector-like coupling of DM to the
vector fields, the contributions corresponding to the longitudinal
polarization modes of the massive gauge bosons disappear
identically. Furthermore, as the momentum exchanged is small
(typically in the range of 10--100 MeV for non-relativistic DM), the
Mandelstam variable $t$ can safely be neglected in all but the photon
diagram. In other words, the amplitude depends on combinations such as
\begin{equation} \label{eq:ampfac}
    - t^{-1} \,  g^{\chi, 1} g^{q, 1}_V + m_Z^{-2} g^{\chi, 0} (g^{q, 0}_V+ g^{q, 0}_A \gamma^5)
+ \sum_{k = 1}^{N} m_k^{-2} \  g^{\chi, k+1} (g^{q, k+1}_V + g^{q, k+1}_A \gamma^5) \ ,
\end{equation}
where we have used the fact that the DM-current is inherently vectorial.
  
As we have seen earlier, the DM's couplings to both the photon and the
$Z$ are highly (and almost equally) suppressed, while those to the
heavier gauge bosons are not. And since $m_Z^2 \gg |t|$, the
$Z$-contribution can be safely neglected, leaving only the photon and
the $Z'$ bosons as relevant mediators. 

For the case of the vector-like part of the quark current, the combination
  of Eq.~\eqref{eq:ampfac} can, thus, be approximated by
\begin{equation} \label{eq:prop2}
\mathcal{P} \approx \mathcal{P}_\gamma + \sum _{k=1}^{N}\mathcal{P}_k \, ,
\end{equation}
 with
 \begin{align} 
     \mathcal{P}_\gamma &= \frac{e^2 Q_q Y_x}{2~q^Nt} \label{eq:prop2gamma} \\
     \mathcal{P}_k  &= \frac{Y_\chi(Y_L^q+Y_R^q)}{N} \frac{q}{f^2} \frac{ (-1)^{k}}{\lambda_{k}^2}\sin^2 \left(\frac{k~\pi} {N+1}\right) \label{eq:prop2k}
 \end{align}
 being the contributions from photon and \(Z'_k\) mediation, respectively.

For the axial quark current, the photon term is, of course, absent. The $Z'$-mediated contributions have the same structure as above, with
$(Y_L^q - Y_R^q)$ replacing $(Y_L^q + Y_R^q)$ and, therefore, exhibits
the same qualitative behavior as in the case of the
vector current.

\begin{figure}[t]
\subfloat[]{\includegraphics[width = 2.8in]{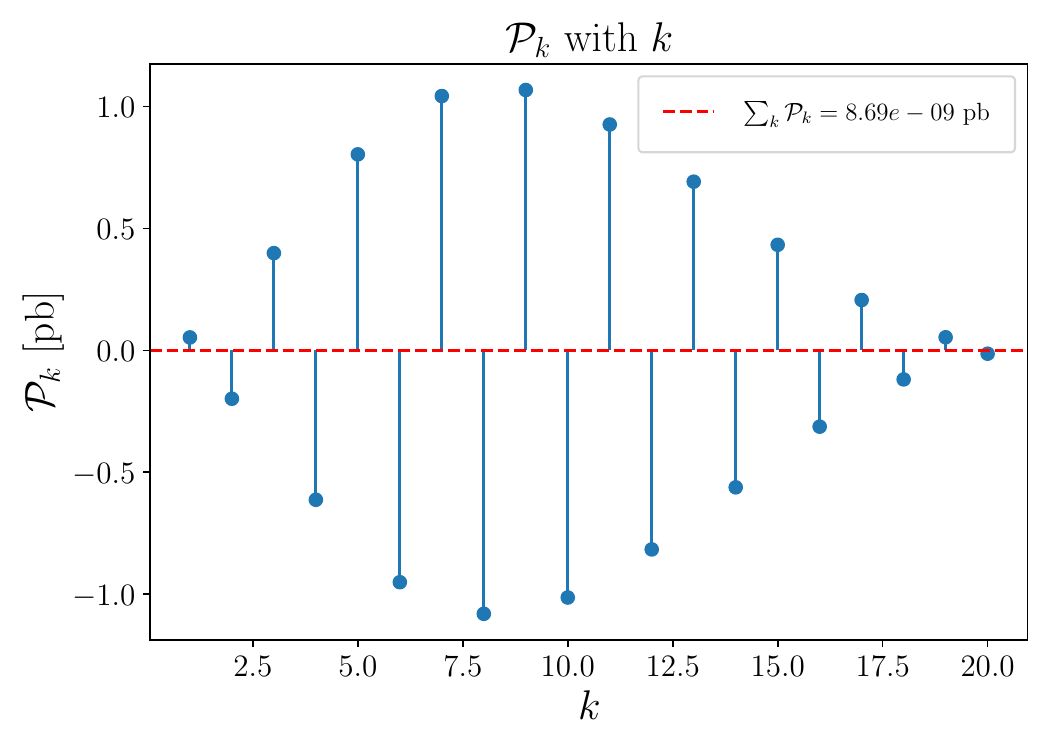}} \quad
\subfloat[]{\includegraphics[width = 2.8in]{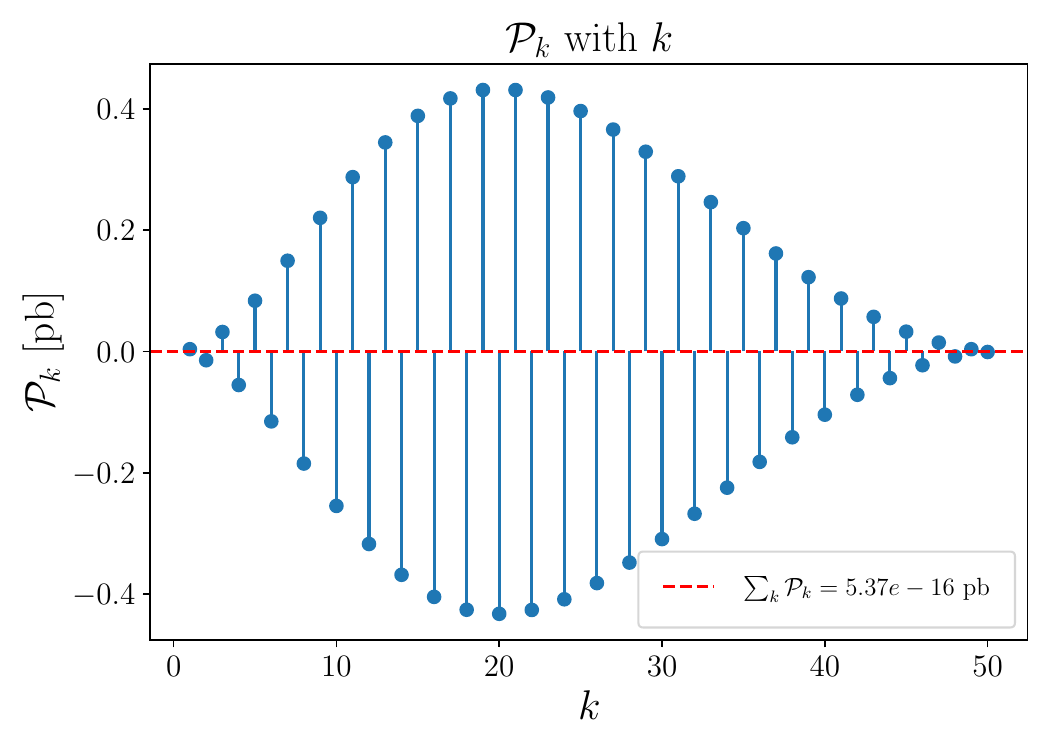}}
\caption{Variation of \(\mathcal{P}_k\) for \(\chi \bar \chi \rightarrow e^- e^+\) ( \emph{i.e} \(Y_L=-0.5, Y_R=-1.0\) and taken \(Y_\chi=1.0\)), given by \eqref{eq:prop2k}  corresponding to different $Z'_k$ states for $q = 3.0$ and $f = 3000~\text{GeV}$, shown for two different cases: (a) $N = 20$ and (b) $N = 50$. The blue dots represent the individual contributions (in picobarns) from each $Z'_k$, while the red line denotes the summed contribution, which exhibits a pronounced cancellation effect.}
\label{plot:amplitudes1}
\end{figure}

At this stage, it behoves us to reexamine the case of all the $Z'$s
being infinitely heavy. The amplitude, then, would be dominated by the
photonic diagram alone. Since a cryogenic bolometer is sensitive to a
small deposition of energy, the detectable recoil cross section can be
sufficiently large unless $g^{\chi, 0}$ is small enough. This is why
the constraint on the charge of the DM is orders of magnitude stronger
than what the name ``millicharged'' conveys.  A second important
feature, characteristic of the clockwork construction, is manifest in
the collective effect of the entire set of $Z'$ bosons. Note that the
individual contributions alternate in sign, leading to significant
cancellations. As a result, the net amplitude is strongly suppressed
despite the individual $Z'$ states having $\mathcal{O}(1)$ couplings
to DM. This behavior is illustrated in Fig.~(\ref{plot:amplitudes1}),
where the total summed contribution is compared against the individual
terms for representative benchmark points. Apart from
this, a further global suppression with increasing $N$ is also
observed if we compare subplots in Fig.~(\ref{plot:amplitudes1}), and
is consistent with the explicit $N^{-1}$ dependence in
Eq.~\eqref{eq:ampfac}. If we compare this to the photon contribution,
which scales as $q^{-N}$, we see that, for sufficiently large
$N$, the photon contribution eventually becomes subdominant, despite
the collective cancellations among the $Z'$ modes. On the other hand,
for moderate values of $N$---the phenomenologically attractive
situation---the photon channel remains dominant by several orders of
magnitude, allowing us to safely neglect the \(Z'\) contributions in
this context.

Having established the relative roles of the photon, $Z$, and $Z'$
channels, we now turn to the phenomenologically relevant region of the
parameter space and the corresponding experimental constraints.  Since
our focus here is on millicharged dark matter, we first consider the
photon-mediated contribution. An estimate of the electromagnetic cross
section for dark matter--proton scattering is given
by~\cite{Iles:2024zka}
\begin{equation}
    \sigma_p=\frac{16\pi \alpha^2\epsilon^2 \mu_{\chi p}^2}{(2m_N E_R)^2},
    \label{eq:sigman}
\end{equation}
where $\sigma_p$ denotes the scattering cross section between the
proton and the dark matter particle $\chi$. Here $\alpha$ is the
fine-structure constant, $\epsilon$ is the charge ratio between the
dark matter particle and the proton, $\mu_{\chi p}$ is the reduced
mass of the proton--$\chi$ system, $m_N$ is the nuclear mass of the
target, and $E_R$ is the recoil energy imparted during
scattering. Since the nucleon charge is simply the sum of the
constituent quark charges, the DM--quark interaction straightforwardly
translates to the DM--proton interaction. If the photon is the only
mediator between the DM and the visible sector, direct detection bounds
constrain the parameters $q$ and $N$ through the relation $\epsilon =
Y_\chi q^{-N}$, with $Y_\chi$ taken to be an $\mathcal{O}(1)$ quantity. From
current direct detection results, the most
stringent limits arise from the LUX-ZEPLIN (LZ-2025) experiment\cite{LZ:2024zvo}. Over a wide
range of dark matter masses, extending from \(\mathcal{O}(10)\,\text{GeV}\) to the multi-TeV scale, LZ sets upper limits on the spin-independent DM--nucleon scattering cross section at the level of \(10^{-12}\!-\!10^{-10}\,\text{pb}\).

Using Eq.~\eqref{eq:sigman} for a Xenon target, this implies
$\epsilon \sim 10^{-12}$ for $\mathcal{O}(1)$ TeV DM. This bound translates into a lower limit on
$N$ for a given $q$. For example, for $q \simeq 3.0$, one requires $N
\gtrsim 24$. The most interesting scenarios are, therefore, those with
relatively small $N$, since these lie within the reach of future
searches.For
($q \simeq 3.0$, $N\simeq25$), the collective $Z'$ contribution is
suppressed by several orders of magnitude compared to the photon
channel, and becomes comparable only for $N \sim 30$. Thus, our
earlier assumption that the heavy vector contributions can be
neglected in such a region remains well justified.

\subsubsection{Collider Bounds}

At the LHC, the $Z'_k$ may, of course, be produced, at the tree order,
from $q\bar q$-fusion.  Of the decay modes into the SM particles, the
dominant one is into a $q \bar q$ pair. However,
the QCD backgrounds to dijet production are overwhelmingly large,
notwithstanding the existence of resonances. Hence, we turn to
the Drell-Yan process, namely $p p \to \ell^+ \ell^-$.
The relevant parameter $R_k$ is defined as~\cite{CMS:2021ctt}
\begin{equation} \label{eq:dilepratio}
    R_k = 
    \frac{\sigma(pp \rightarrow Z'_k \rightarrow \ell^- \ell^+)}
         {\sigma(pp \rightarrow Z \rightarrow \ell^- \ell^+)}
         \longrightarrow
         \frac{\sigma(pp \rightarrow Z'_k)}{\sigma(pp \rightarrow Z)} \,
          \times \,
         \frac{\mathcal{BR}(Z' \rightarrow \ell^- \ell^+)}
         { \mathcal{BR}(Z \rightarrow \ell^- \ell^+)} \, ,
\end{equation}
with the second step holding under the narrow width approximation.
The first fraction on the right-hand side (resonant production) can be easily calculated, even incorporating the QCD-corrections as available in the literature~\cite{Hamberg:1990np}, or even estimated from existing $Z'$ studies by accordingly scaling with the couplings. As for the second factor (ratio of branching fractions), couplings of a given $Z'_k$ to the SM fermions are proportional to their hypercharges, and thus, $\Gamma(Z'_k \to \ell^+ \ell^-) / \Gamma(Z'_k \to SM)$ is expected to be approximately SM-like (the opening of the $t 
\Bar{t}$ channel does not significantly alter this ratio).  Nonetheless, note that one of the major decay modes of the $Z'_k$ is that in the DM {\em viz.}, $\chi\bar\chi$. However, as it would turn out, the reproduction of the correct relic abundance would stipulate $Y_\chi \lesssim 0.5$ leading to $\mathrm{BR}(Z_k' \to \chi\bar{\chi}) \lesssim 0.4$. The presence of this additional channel, though, does not render the aforementioned ratio of the branching significantly smaller than order unity.
\begin{figure}[htb]
    \centering
    \includegraphics[width=0.76\linewidth,height=0.46\linewidth]{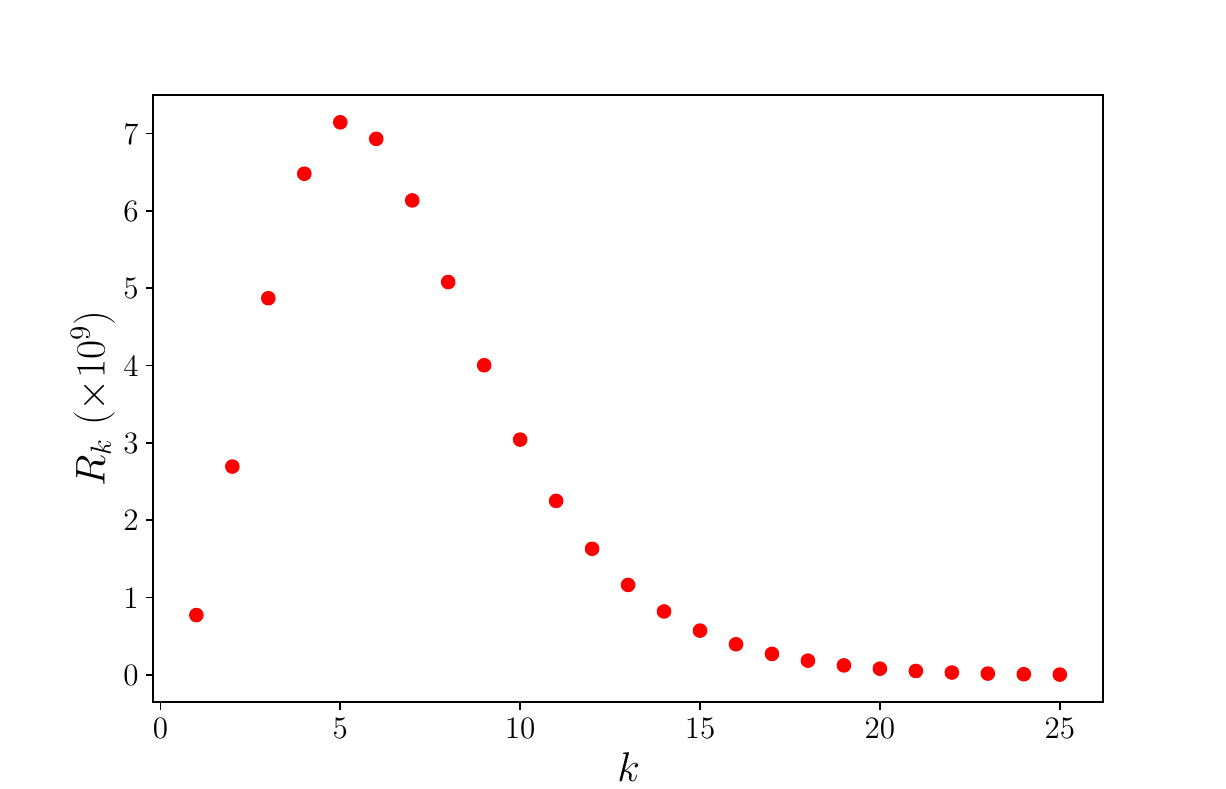}
    \caption{Plot of $R_k$ for the individual $Z'_k$ modes,
        where $R_k$ denotes the dilepton production rate mediated
        by each $Z'_k$ relative to the $Z$ boson.}
    \label{fig:r_sig}
\end{figure}
\begin{figure}[!h]
    \centering
    \includegraphics[width=0.76\linewidth,height=0.48\linewidth]{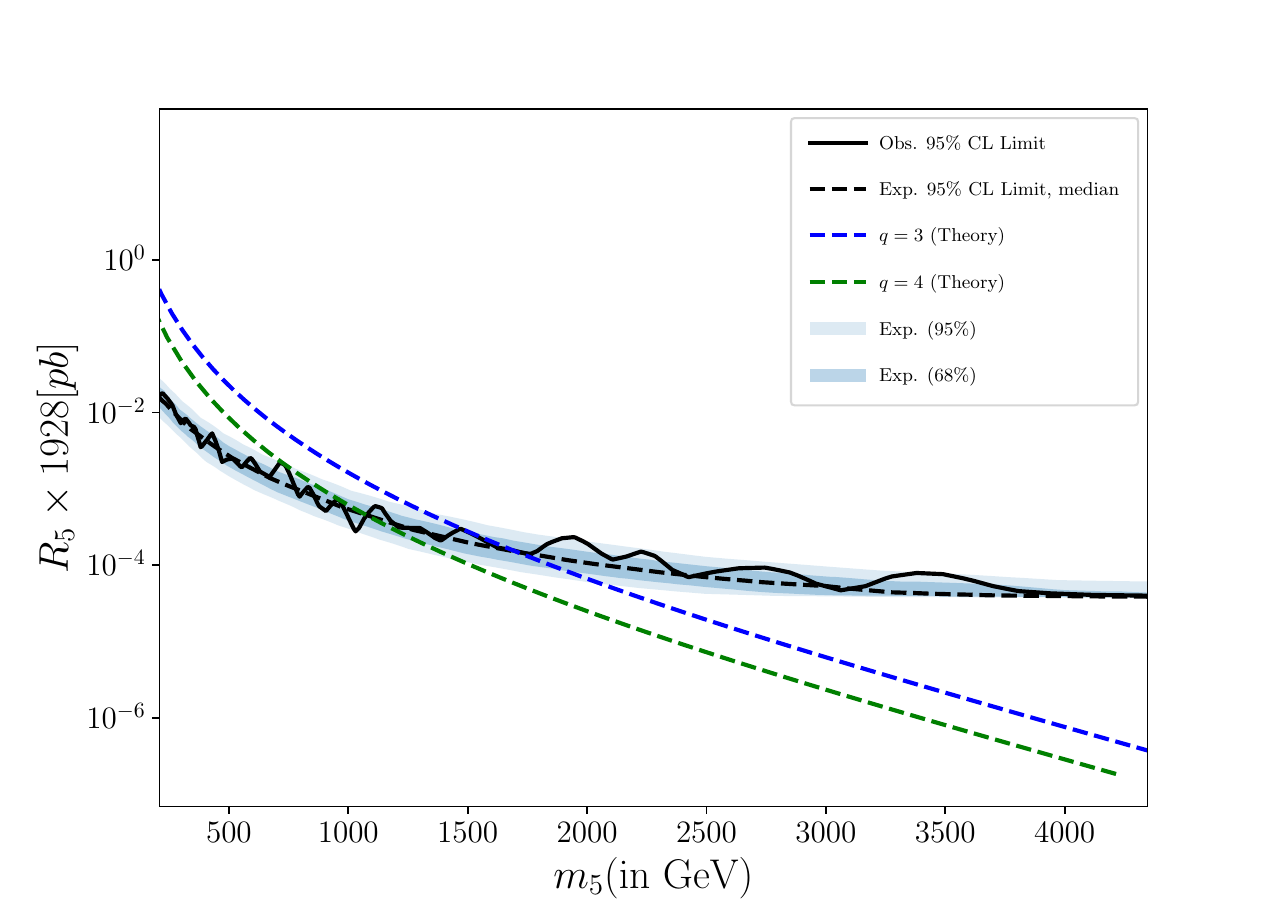}
    \caption{Comparison of $R_5$ (scaled by $1928~\mathrm{pb}$), as defined in Eq.~\eqref{eq:dilepratio}, for two benchmark values $q=3$ (blue dashed curve) and $q=4$ (green dashed curve) for $N=25$, with the 95\% CL observed and expected limits from CMS dilepton searches at $\sqrt{s}=13~\mathrm{TeV}$ and $L=137~\mathrm{fb}^{-1}$\cite{CMS:2021ctt}. The blue bands denote the uncertainty, with the darker (lighter) shaded region corresponding to the 68\% (95\%) confidence level.}
    \label{fig:r5_dilep}
\end{figure}

Going beyond the naive estimates, we present, in
Fig.~(\ref{fig:r_sig}), the values of $R_k$, for a specific benchmark
point that would be seen (in next section) to yield the correct relic
density, \emph{i.e.} \(q = 3, \ N = 25, \ f = 3~\text{TeV}, \ Y_\chi =
0.2\). The largest relative signal strength, in this case, occurs for
$k = 5$ (with $m_5 \simeq 2564~\text{GeV}$). Therefore, a
representative lower bound on the SSB scale $f$ can be obtained by
comparing the relative signal strengths in the model with the
corresponding experimental limits from dilepton searches
\cite{CMS:2021ctt,ATLAS:2017fih} for varying masses of $Z'_5$. This is
carried out in Fig.~(\ref{fig:r5_dilep}) for $q=3,4$, which translates
into the lower bounds $f \gtrsim 2$ TeV and $f \gtrsim 1$ TeV, respectively. A lower value of \(f\) can yet be achieved for a larger \(q\), as provisioned by \(m_{Z'} \sim g_x q f\). However, since $g_x q \gg 1$ can potentially come in conflict with theoretical issues pertaining to the RG evolution of the scalar potential (see sec.(\ref{subsec:theocons})), the effective lower limit on the scale \(f\) remains largely comparable to that on $m_{Z'}$ itself.

Although the aforementioned limits have been derived for the case of $N=25$, it can be argued that these will be effectively valid for any $N$ as the mode ($k$) yielding the largest signal strength will also accordingly vary.

At this stage, we turn to the dominant decay mode, namely $Z'_k \to \chi \bar\chi$. Since the $\chi$ is invisible, one must, instead, turn to a visible particle in the final state so as to trigger the detector.  The lowest order processes leading to such a signal are $q + g \to \chi + \bar\chi + q $ or $q + \bar q \to \chi + \bar\chi + g/\gamma $. The corresponding final state comprises a monojet (or monophoton) accompanied by missing transverse momentum, and is well-studied in the context of the LHC. Once again, the expected cross-sections are below the current sensitivities.

\section{Dark matter phenomenology} \label{sec:DMpheno}

Now that the model has been completely defined and its consistency
with the standard EW physics established in terms of relevant
constraints on its parameters, we have the requisite foundation to
describe, in detail, the phenomenology of the DM candidate
$\chi$. It is clear from Eqs.~\eqref{eq:lag2} and
(\ref{eq:covchi}) that the interactions of $\chi$ with the SM fermions
are mediated by the photon, the $Z$ boson, and the tower of heavy
$Z'$s. As we have argued in Sec.~(\ref{sec:cons}), weak-scale CHAMPs must
have only very feeble couplings with the photon and $Z$ so as to be
consistent with limits from the direct detection experiments, and this
is naturally achieved via the clockwork mechanism. The primary
mediators, then, are the $Z'$s that have unsuppressed (periodic) gauge
couplings with $\chi$ as well as the SM fermions. As a result, the DM
relic abundance in our setup is determined predominantly through the
freeze-out mechanism. This entails the CHAMP decoupling from the SM
bath once the temperature falls well below $T\sim m_{\chi}$ and
eventually freezing-out as its rate of annihilation into the visible
sector becomes comparable with that of the Hubble
expansion\footnote{If the cross sections are comparable to weak
interaction ones, the freeze-out is expected to occur at a temperature
$T_{fo}\sim m_{\chi}/20$.}.

Certain interesting phenomenological features characteristic to the
clockwork portal emerge upon a careful investigation of the
annihilation dynamics of the DM. This is best illustrated by assuming
two distinct scenarios pertaining to the primary annihilation channels
$\chi \Bar{\chi} \to \psi \Bar{\psi}$, $\chi \Bar{\chi} \to Z'_k
Z'_k$, and $\chi \Bar{\chi} \to Z'_k \phi_{n}$, where $\psi$ denotes
the SM fermions. In the following, we first discuss in detail the case
wherein only the SM channels are kinematically accessible and,
thereafter, describe the more general case, including all possible
channels.

\subsection{The simplest scenario: $\chi \Bar{\chi} \to \psi \Bar{\psi}$}

Annihilations to the SM particles dominate when $m_k, m_{\phi} \gtrsim
2m_{\chi}$ and the $H-\Phi_n$ mixings are small. Even within the SM
channels, the fermionic final states dominate over the rest, namely
the $W^+ W^-$, $ZZ$ and $Zh$ channels. This can be simply understood
from the fact that the $Z'WW$
couplings are induced through the $B^{k}-W^3$ mixings (post CWSB)
and hence carry a $v^2/{q^4 f^2}$ suppression. Similarly, the subdominant nature of the $Zh$ channel can be
attributed to the $Z'Zh$ couplings that are suppressed by both a
$v^2/f^2$ factor and the scalar mixing angle $\sin \zeta$. On the
other hand, the $t$-channel annihilation to $ZZ$ is trivially
suppressed by the $q^{-N}$ factor in the $Z\Bar{\chi} \chi$ coupling.

The key difference between a generic $Z'$ portal and the present case
is manifested in the range of allowed DM masses consistent with the
observed relic abundance. For a typical case of a $Z'$-portal, two
qualitatively different situations prevail. Given that the DM is a
cold gas, if $m_{Z'} \gapp 2 m_\chi$, a resonance enhancement of the
cross section (scaling as $g^2_{Z'}/m^2_{Z'}$) obtains and DM
annihilation is rendered very efficient (occasionally, overly so). On
the other hand, for a $Z'$ far heavier (often the case in various
scenarios), the cross section, away from the resonance, would scale as
$g^4_{Z'} m_\chi^2 / m_{Z'}^4$. Reproducing the observed relic
abundance, then, results, in either case, in a straightforward
constraint in the parameter space.  The said constraint, though, has
to be examined against the bounds from direct (as well as indirect)
searches.

The present case is more complicated.  For one, unlike in canonical
$Z'$ models, the direct detection amplitudes here receives potentially
significant contributions from photon exchanges, thereby affecting the
said constraints. More importantly, we now have multiple $Z'$s, all
relatively closely spaced, with all the couplings determined by a
single gauge coupling $g_x$. With $g_x$, in turn, being close to the
SM hypercharge coupling---see Eq.~\eqref{eq:cons1})---the relic
abundance is primarily determined by three parameters, {\em viz.} the
DM mass, its charge $Y_\chi$ under $U(1)_N$ and the scale $f$ (for it
determines the $Z'$ mass scale). There is also a residual dependence
($q, N)$ through their influence on the masses of the $Z'$s as well as their
individual couplings to the fermions. 

Close to the individual peaks ($m_\chi \sim m_k/2$), the corresponding
amplitude suffers a large resonance enhancement, rendering the other
contributions largely irrelevant. As with the single-$Z'$ models, very
close to resonance, the enhancement could be too strong thereby
suppressing the relic abundance. Naturally, for two values of $m_\chi$
around $m_k/2$ the abundance would be just right. The exact distances
from the central value would depend on the effective individual
couplings of the $Z'$ under question and, hence, its width. Away from
the resonances, the situation is more complicated as the interference
between the individual contributions become increasingly
important. Two effects are in play here. As discussed in the preceding
section, the couplings of the $Z'_k$ oscillate with $k$, leading to
potentially large cancellations between the amplitudes mediated by
them. A further complication is that, unlike in the case of direct
detection as discussed earlier, DM annihilation proceeds through
$s$-channel diagrams and for $m_k^2 < s < m_{k+1}^2$ the two
propagators have opposite signs. In other words, there is partial
cancellation between diagrams with $m_k^2 < s$ owing to the
oscillating nature of the couplings\footnote{These are
  qualitatively similar to the $t$-channel cancellations shown in
  Sec.~(\ref{ssec:ddc}).}  And similar is the case for
diagrams with $m_k^2 > s$. Finally, there is a further cancellation
between the two subsets on account of the relative signs of the two
sets of propagators.

Thus, effectively, the only contribution that matters in this case is
the resonant annihilation of $\chi$ near any one of the $Z'$
thresholds. The process has a dominant $s$-wave component and, hence, the corresponding thermally-averaged cross-section is, to the leading order, independent of $x^{-1}\equiv T/m_\chi$.
Taking any one of the $Z'_k$ as the mediator, and working in the limit where the light SM fermions are effectively massless (an excellent approximation), this is given by 
\begin{equation}
    \langle \sigma v \rangle_{CM} \approx \frac{(g^{\chi,k+1})^2  m_\chi^2\left((g^{f,k+1}_V)^2+(g^{f,k+1}_A)^2\right)}{ \pi \left( 16m_\chi^4-8m_\chi^2 m_k^2 +m_k^4 +m_k^2 \Gamma_k^2\right)} \, ,
\end{equation}
where the various couplings are defined in Sec.~(\ref{sec:mod}) and $\Gamma_k$ is the decay width of the $Z'_k$.  Note that since the
$Z'$s and, hence, the heavy scalars $\phi$s, are phenomenologically
required to have masses $\gtrsim 1$ TeV, they will generally decouple
from the thermal bath pretty early and then promptly decay into the SM
fermions.

To illustrate the aforementioned dynamics with an example, we choose a
configuration such that the CHAMP is potentially sensitive to the
forthcoming DM direct detection experiments. This simply requires that the DM-nucleon cross-section be close to the current experimental limits for a given DM mass. Since collider constraints already imply $m_{Z'} \gtrsim 1$ TeV, we accordingly assume, for resonant annihilation, DM masses around similar scales. In compliance with the direct detection bounds we choose the parameter configuration \(N = 25\), \(q = 3.0\), and
\(f = 3~\text{TeV}\), for $Y_{\chi}=0.2$, in which case the
$Z'$ spectrum appears with masses ranging from \(2.2 - 4.6
\ \text{TeV}\).
\begin{figure}
     \centering
     \includegraphics[width=0.8\linewidth]{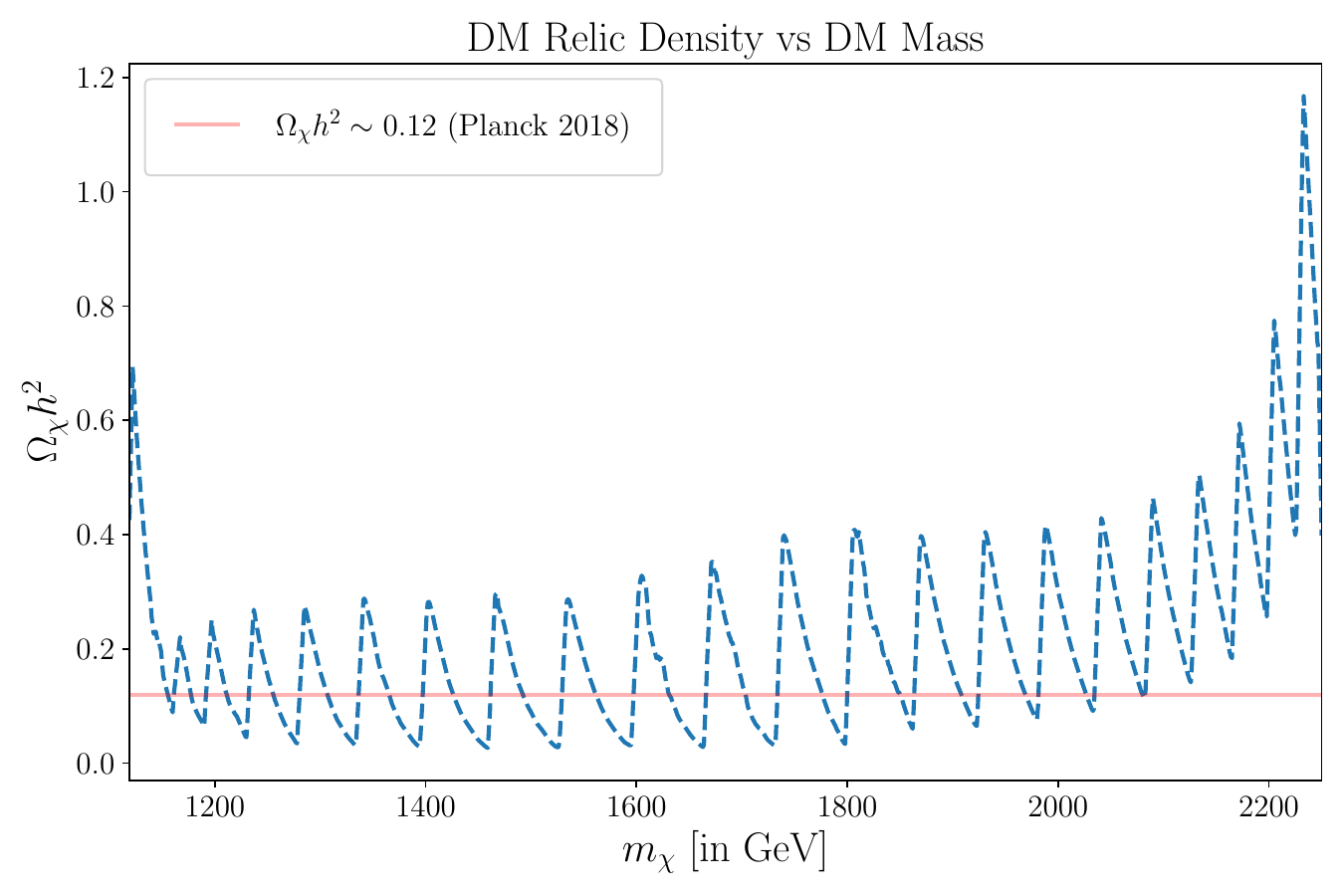}
     \caption{DM relic density (shown in dashed blue) for different values of its mass (as calculated with \texttt{MadDMv.3.2}~\cite{Backovic:2013dpa,Backovic:2015cra}) for the benchmark $N = 25$, $q = 3.0$, $f = 3~\text{TeV}$ and $Y_\chi =
       0.2$. The red line indicates the reference value of the observed relic abundance, $\Omega_\chi h^2 =       0.120 \pm 0.001$~\cite{Planck:2018vyg}. Here, we focus on the most relevant region and show the variation only over a finite range of the DM mass where resonant annihilations through the $Z'$s dictate the DM yield, \emph{i.e.} $m_{\chi} \in (m_1/2, \, m_N/2)$.}
     \label{fig:relic_bm1}
 \end{figure}
 The correct relic abundance for \(\chi\) can be achieved in this
 setup in the vicinity of the resonance regions, as shown in
 Fig.~(\ref{fig:relic_bm1}). Away from the resonances, the annihilation
 cross section is strongly suppressed due to the destructive
 interference in the total amplitude characteristic of the clockwork
 configuration, as discussed earlier, which leads to an overabundance
 of $\chi$. However, resonance enhancement kicks in as the dark matter
 mass, \(m_\chi\), approaches the band of the CW $Z'$s, which results
 in a significant increase in the annihilation cross section. This, in
 turn, results in a sizable decline in the relic
 density. Consequently, the relic density curve exhibits a
 characteristic wavy pattern along the $m_{\chi}$ axis, which
 periodically coincides with the line corresponding to the correct
 relic abundance (shown in green), as seen in
 Fig.~(\ref{fig:relic_bm1}).

Looking more closely, the asymmetric shape of the relic density
profile across the individual resonances, as seen in
Fig.~(\ref{fig:relic_bm1}), can be well understood within the narrow
width approximation (NWA). In the NWA the thermally averaged
annihilation cross section, in the limit that the SM fermion masses
are negligible, can be approximated as \cite{GONDOLO1991145}
\begin{align} \label{eq:sigvnrw}
    \langle \sigma v \rangle &=  \frac{1}{96 \ m_\chi^4 T K_2^2(m_\chi/T)} 
    \sum_{k=1}^{N}  (g^{\chi,k+1})^2 \left((g^{f,k+1}_V)^2 + (g^{f,k+1}_A)^2 \right) \nonumber \\ & \times \frac{m_k\left(m_k^2+ 2 m_\chi^2 \right)}{\Gamma_k} \sqrt{\left(m_k^2-4m_\chi^2\right)} \ K_1\left(\frac{m_k}{T} \right) ~ \theta(m_k-2m_\chi) \, ,
\end{align}
where \(K_{1,2}\) are the modified Bessel functions of the first and
the second kind, respectively. The Heaviside function
rejects the contribution to $\langle\sigma v \rangle$ mediated by a
particular $Z'_k$ if the DM mass exceeds $m_k/2$. This, of
  course, is only an approximation as an off-shell $Z'_k$ would
  contribute too. However, given the discussion above, it is easy to
  see that the cross section falls very fast away from the
  resonance. Consequently, for $m_\chi >
m_k/2$, the contribution from that particular resonance may be
  well approximated to be vanishingly small since the
minimum center-of-mass energy $\sqrt{s} = 2 m_\chi$ exceeds $m_k$,
making it impossible to hit the resonance.  This inevitably results in
a sharp rise of the relic density profile on one side of the
resonance.

In contrast, for $m_\chi < m_k/2$, annihilation through the resonance
still remains viable. In this case, the thermal motion of DM particles
allows them to collide with sufficient energy to access the resonance
region. This effect is captured by the thermal averaging procedure,
where the momentum distribution of DM particles, encoded in the Bessel
functions appearing in Eq.~\eqref{eq:sigvnrw}, ensures that a
fraction of collisions occur at or near the resonance, thereby
enhancing the annihilation cross section. This particular effect
corresponds to the regions in the profile with a smooth variation near
the individual resonances, as shown in Fig.~(\ref{fig:relic_bm1}).

\subsection{Including all annihilation channels}

In our discussions thus far, we have limited ourselves to DM
annihilations into SM particles alone. This, of course, leads to an
exact result as long as annihilations into the new particles are
kinematically forbidden\footnote{Note that with the $\chi$ having only
gauge couplings, co-annihilation processes are not possible.}. We now
examine the consequences when this assumption no longer holds and
other channels might be important too.  This assumes importance as,
for sufficiently large $m_\chi$, the annihilation into the SM
particles is no longer efficient enough to suppress the relic
abundance down to the required level.

With the $\gamma$ and $Z$ couplings of the DM being very small,
processes such as $\bar \chi + \chi \to Z'_k + Z/\gamma$ are of little
interest, and the only relevant additional channels are those to $Z'_k
+ Z'_l$ and $Z'_k + \varphi_n$ respectively. We examine these in
turn, and we only discuss the salient features here.

\subsubsection{\(\chi \bar{\chi} \to Z'_kZ'_l\)}

 Kinematically accessible only when \(2m_\chi > m_{Z'_k} + m_{Z'_l}\),
 these proceed through \(t\)-channel exchanges of \(\chi\)
 itself. With a plethora of final states being available (at least for
 a sufficiently heavy $\chi$), it might seem that the total
 annihilation cross section can be substantial. And with the $Z'$s
 decaying promptly, this would be expected to suppress the relic
 abundance to a significant degree. However, it should be realised
 that couplings of the $\chi$ to the $Z's$ all descend from a single
 one, {\em viz.} the $\bar \chi \chi X_N$ term and,
 in the unbroken phase, there would exist only the single process
 $\bar \chi \chi \to X_N + X_N$. Thus, in the limit of equal $Z'$
 masses,
  \[
  \sum_{ij} \sigma (\chi \bar{\chi} \to Z'_kZ'_l) \simeq
      \sigma(\chi \bar{\chi} \to X_N X_N) \ ,
\]
with the right hand side being evaluated in the absence of gauge boson
mixing.  In the above, the equality would be exact if the sum were to
include the photon and the $Z$ and, thus, the discrepancy is tiny
indeed. The limit of equal $Z'$-masses, of course, does not hold. However,
with the relative splittings being small, the above still continues to be a
good approximation.
      
For the parameters chosen above, this extra contribution to DM
annihilation is not sufficient. Indeed, with $g_x$ being constrained
to be very close to the hypercharge coupling within the SM, the $\bar
\chi \chi \to X_N + X_N$ process can be enhanced only by either
enhancing the $U(1)_N$ charge $Y_\chi$ or by reducing the $Z'$
masses. The former alternative has an obvious limit from the
requirement of perturbativity. More crucially, though, enhancing
$Y_\chi$ indiscriminately would also increase the DM coupling with the
photon, thereby coming into conflict with direct detection
experiments. This, in turn, could be compensated for by increasing
either $q$ or $N$, with these being subject to various theoretical and
experimental bounds as mentioned in Sec.~(\ref{sec:cons}). The second
option, {\em viz.} making the $Z'$s lighter, could, in principle, be
achieved by lowering the clockwork scale $f$. However, \(f \lesssim
1~\text{TeV}\) is disfavoured from negative results for the dilepton searches. 

It must be noted, though, that while the $Z'Z'$ channel, by itself, is
largely insufficient in yielding the correct DM abundance for a large
portion of the parameter space, it is quite possible that other
channels compensate for the required cross-section. The contributions
of the remaining channels to this effect are examined in the
following.

\subsubsection{\(\chi \bar{\chi} \to  Z'_k \varphi_n, Z'_k \Phi\)}

With all the putative pseudoscalars being absorbed by the gauge
bosons, the only remaining possibilities are the Bjorken-process
analogues ({\em viz.}, $\chi \bar{\chi} \to Z'_k \varphi_n, Z'_k
\Phi$) that open up for \(2m_\chi \gtrsim m_k + m_{\varphi}\) and
\(2m_\chi \gtrsim m_k + m_\Phi\) respectively.  The scalar masses---see Eq.~\eqref{eq:eigvheavyscal}---are governed by the two essentially
free parameters $\xi$ and $\eta$ in the potential---see
Eq.~\eqref{eq:pot}---with the only constraint being observability at
the LHC (as highlighted in Sec.~(\ref{sec:cons})).

Once again, the plethora of possible final states is misleading as,
above the clockwork scale, the only relevant annihilation amplitude is
$\chi \bar{\chi} \to X_N \phi_{N-1}$ and all the amplitudes discussed
above descend from this single one. In terms of the mass eigenstates
($\varphi_n, \Phi$), nondiagonal vertices $Z'_k Z'_l \varphi_n$ {\em
  etc.} exist and the individual annihilation processes are, in
principle, mediated by all the $Z's$. Consequently, an analogue of the
destructive inference operative for the fermionic channel also holds
here, albeit with the cancellations becoming less significant as $k
\to N$.

Nevertheless, the $Z'\varphi$ annihilation channels, combined with the
$Z'Z'$ ones, can indeed yield the typical WIMP DM cross-section
$\langle \sigma v\rangle_{\mbox{ann}} \sim 10^{-26} \,
\mbox{cm}^3/\mbox{s}$, provided $m_{Z'}, m_{\varphi}\lesssim 1$
TeV. Since the latter requirement invariably translates to $f \lesssim
1$ TeV, only a small region of the parameter space remains viable for
a DM candidate, consistent with the other constraints. For instance,
one such allowed configuration goes as --- $m_{\chi} \sim 1.5$ TeV,
$Y_{\chi} \sim \mathcal{O}(1)$, $f\sim 1$ TeV ($m_k^{\mbox{max}}\sim
2$ TeV), $m_{\varphi} \sim 1$ TeV, $q=4$ and $N \sim 20$.

In conclusion, it is perhaps worth drawing a qualitative comparison of
the nonresonant annihilation to $Z'/\varphi$ with the resonant
annihilation to the SM fermions. Since the success of the combination
of the $Z'_kZ'_l$ and $Z'_k \varphi_n$ channels in setting the correct
DM abundance is contingent on the $Z'$ and $\varphi$ masses being
close to the current experimental bounds, and having $U(1)$ charges
near perturbative thresholds, the scenario is imminently and
independently falsifiable from future electroweak precision
measurements as well as dedicated $Z'$ and heavy Higgs searches. In
contrast, the $f \Bar{f}$ channel becomes relevant only at the $Z'_k$
resonances and, hence, is very characteristic of the clockwork
portal. Thus, even if the next generation of direct-detection
experiments place stricter limits on the DM-nucleon cross-section, or
future $Z'$ searches push the lower limit on their masses beyond the
multi-TeV range, the resonance portal would still allow for a natural
realization of a CHAMP DM as it only requires a modification in $q$
and $N$, alongwith with the assumption little hierarchy between $v$
and $f$.

\section{Summary and Discussion} 
\label{sec:disc}

We revisit the possibility of an electrically charged dark matter
candidate for the highly constrained, and hitherto less-explored, case
of a near weak-scale charged massive particle (CHAMP). To realize a
scenario with no unnaturally small parameters, we invoke a gauged
\emph{clockwork} portal mediating the dark and the visible sectors,
which not only helps generate the required suppression in the electric
charge of the DM but also serves to determine the correct DM relic
abundance through a freeze-out mechanism. The clockwork sector
comprises $(N+1)$ copies of an Abelian gauge field, with $N$ complex
scalars suitably charged so as to act as \emph{links} between adjacent
gauge fields, exhibiting a lattice structure with only
nearest-neighbour interactions. Resorting to the most minimal picture,
we invoke this sector as an extension of the SM hypercharge ($U(1)_Y
\to U(1)^{N+1}$) such that the zero mode of the spectrum, post the spontaneous breaking of
  the clockwork sector symmetry, emulates the SM boson $B_{\mu}$. In
the physical basis, then, this results in the photon being intertwined
with the CW sector---rather than a purely SM entity---with an
exponentially falling distribution over the CW lattice. With the SM
particles charged under $U(1)_0$ and the CHAMP charged under $U(1)_N$,
this leads to the photon having the usual $\mathcal{O}(1)$ couplings
with the SM particles and a $q^{-N}$-suppressed interaction with the
DM. Thus, for $q>1$, the DM can be made
to effectively carry a very tiny electric charge, even if we assign an
$\mathcal{O}(1)$ charge to it under $U(1)_N$.

For simplicity, and to avoid complications related to anomalies, we
assume the DM field ($\chi$) to be a single vector-like Dirac fermion,
which trivially renders it stable. It interacts
with the visible sector dominantly through the $N$ heavy $Z'$s present
in the CW spectrum (with the couplings being mandated by the gauge
boson mixings engendered by the CW mechanism), whose masses are
determined by the SSB scale $f(\gg v)$. The DM abundance is then set
by a conventional freeze-out mechanism, dictated by DM annihilation to
the SM particles. To this effect, the most accessible channel in terms
of the available parameter space turns out to be the $Z'$ mediated
annihilation to the SM fermions. The correct relic density is reached
only when $m_{\chi} \sim m_{Z'}/2$, for any $Z'$, since away from the
poles the individual $Z'$-mediated contributions destructively
interfere and undergo severe cancellations. This sets a characteristic
scale for the DM mass in the model, inevitably tied to the symmetry
breaking scale $f$.

Although, naively, a setup as such might seem \emph{ad hoc} and
overtly complicated in nature, it is important to note that it can
possibly be viewed as a deconstruction of a five-dimensional
braneworld scenario wherein the hypercharge gauge boson is placed in
the bulk. In that case, the rest of the SM fields and the DM should be
localized on the IR and the UV branes, respectively. The relevant
geometry, as suggested in ref.\cite{Giudice:2016yja}, can be generated
by a linear dilaton augmentation of 5D gravity
\cite{Antoniadis:2011qw,Cox:2012ee}.

The most important theoretical constraint on the setup is due to the
scalar potential stability and reads $N \eta^2 < 4 \lambda \xi$.  However,
phenomenologically, the most relevant bounds on the model arise from
experimental observations. For instance, the current upper limit on
the DM-nucleon scattering cross-section, from LUX-ZEPLIN places the
most stringent upper bound on the electric charge of the DM (\(\sim
10^{-12}\)). Since even for the $t$-channel processes the individual
$Z'$ contributions effectively cancel out due to the periodic nature
of the couplings, the DM-nucleon scattering in the model proceeds
primarily through a photon exchange and, therefore, the aforementioned
bound directly limits the combination $q^{-N}$. On the other hand, negative results for dilepton searches translate to a
lower bound on the $Z'$ mass scale, namely $m_Z'
\gtrsim \mathcal{O}(1)$ TeV, which happens to be stronger than the
bounds derived from direct searches at the LHC (for effective couplings as applicable to this model).

The clockwork portal scenario for a CHAMP stands out in two major
aspects. Firstly, unlike in typical WIMP models, demanding consistency
with persistent null results at the direct detection experiments in
the future will not necessarily render the model unnatural. This is
simply because the required suppression in the rates can always be
accommodated by adjusting the number of portal fields $N$ accordingly
(and perhaps also marginally increasing the hopping charge $q$ within
the perturbativity limits). While this might seem uninteresting in
terms of the testability of the model, it nevertheless exhibits
robustness against the potentially tightening constraints from the
next generation of direct search experiments.  Secondly, since the
most relevant case of a viable DM involves resonant annihilations to
SM fermions, the masses of the heavy mediators (\emph{i.e.} the $Z'$s)
are neither constrained from the required DM yield, nor from the
stringent direct detection limits as the DM-nucleon scatterings are
dominated by photon exchanges alone. Therefore, a plausible CHAMP
scenario, with TeV scale masses for both the DM and the heavy
mediators, seems quite straightforwardly achievable
in this setup. There is a caveat, though, to both the arguments
  above.  While increasing $q^N$ does suppress the photon exchange
  contribution to the direct detection cross sections, the heavy gauge
  mediated contributions remain largely unaffected and, thus, future
  experiments would still be able to probe the model, albeit in a more
  traditional mode. Moreover, the increased $q^N$ suppression has
  little effect on the electroweak precision
  observables. Consequently, the model would remain eminently testable
  at a FCC-ee machine, with a projected goal of producing  $\gtrsim 10^{12}$ $Z$-bosons.
  
Since the DM in the setup favours annihilations into a pair of charged
fermions through a $Z'$, it is potentially sensitive to an array of
forthcoming (or planned) MeV scale $\gamma$-ray telescopes such as
\texttt{AMEGO}~\cite{AMEGO:2019gny,2021SPIE11444E..31K,Caputo:2022xpx},
\texttt{E-ASTROGAM}~\cite{e-ASTROGAM:2016bph,e-ASTROGAM:2017pxr} and
\texttt{MAST}~\cite{Dzhatdoev:2019kay}, which are expected to target
weak-scale DM candidates as well. Even without a dedicated study of
the $\gamma$-ray fluxes within the model, a qualitative inference can
be sought from the model-independent analysis in
ref.\cite{Cirelli:2025qxx} of the sensitivity projections of the
aforementioned experiments for a weak-scale DM. Of course, a direct or
indirect signal of a weak-scale DM would not necessarily reveal the
nature of the portal as well. Conservatively, then, only a direct
search for the multi-$Z'$ (and the accompanying multi-scalar) spectrum
can qualify as a probe of the clockwork portal. Given the allowed mass
scale of the $Z'$s in the TeV range or beyond, some of the proposed or
planned collider experiments in the high-luminosity (\emph{e.g.}
HL-LHC) or the high-energy frontiers (\emph{e.g.} FCC-hh, multi-TeV
muon collider, etc.) would serve as ideal discovery grounds for such a
scenario.

While we have assumed the popular freeze-out scenario in our work, it
behoves us to consider the alternate, namely the \emph{freeze-in}
mechanism, wherein the DM is expected to be light (sub-GeV), and
produced from either the annihilation or decay of heavier particles
(whether SM or those from the CW sector). With the DM interacting, in
the gauge basis, only with $X_N$ (a coupling distributed, post-mixing,
amongst the $Z'$s), clearly annihilation from a SM pair would be very
suppressed.

In a typical reheating scenario with \(T_{\rm reh} \gtrsim 10^9
\, \mathrm{GeV}\), the SM annihilation channel (\(\rm SM\, SM \to
\bar{\chi}\chi\)) is highly suppressed --- as a consequence of CW localization in the case of $Z Z \to \Bar{\chi} \chi$ or due to significant
cancellations among the $Z'$-mediated contributions away from the poles in the processes $\Bar{f} f \to \Bar{\chi} \chi$, as discussed in Sec.~(\ref{sec:cons}). However, for the parameter space
under consideration, the $Z'$s thermalize with the plasma at high temperatures. As a result, they subsequently decay and annihilate into
dark matter (\(Z' \to \bar{\chi}\chi\), \(Z' Z' \to \bar{\chi}\chi\)), leading generically to an overproduction of dark matter and hence an
overabundant relic density. This behaviour violates the basic
requirement of the freeze-in mechanism, which relies on the mediator
remaining out of thermal equilibrium. Therefore, within the standard
cosmological evolution, the freeze-in scenario is not viable in this
framework. On the other hand, in a non-standard cosmological history,
a low reheating temperature scenario (\(T_{\rm reh} \ll m_{Z'}\)) can
significantly suppress the production of the CW sector. In such a
case, the abundance of CW particles is sufficiently suppressed so that
their contribution, together with the SM-induced production, may yield
the correct dark matter relic density. A detailed study of such
non-standard cosmological scenarios is beyond the scope of the present
work and will be explored in future works.

\section*{Acknowledgement}
DC acknowledges the ANRF, Government of India, for support through the
project CRG/ 2023/008234 and the IoE, University of Delhi grant
IoE/2025-26/12/FRP. V.K.J. acknowledges financial support provided by the UGC, Government of India. S.M. would like to thank Prolay K. Chanda for
helpful discussions. S.M. also acknowledges support from the
Department of Atomic Energy (Government of India).
\newpage
\bibliography{references}
\end{document}